\documentclass{aa} 

\usepackage{graphicx}
\usepackage{txfonts}
\usepackage{natbib}
\bibpunct{(}{)}{;}{a}{}{,}

\newcommand{\be}{\begin{equation}}
\newcommand{\ee}{\end{equation}}
\newcommand{\angstrom}{\mbox{\normalfont\AA}}

\def\apjl{ApJL}
\def\apjs{ApJS}

\newcommand{\Mpc}{$h^{-1}$\thinspace Mpc}

\newcommand{\vmh}{h^{-1}\mathrm{Mpc} }

\begin{document}  

\title{
Infalling groups and galaxy transformations in the cluster A2142 
}

\author {Maret~Einasto\inst{1} 
\and Boris~Deshev\inst{1,2}
\and Heidi~Lietzen\inst{1}
\and Rain~Kipper\inst{1} 
\and Elmo~Tempel\inst{1,3} 
\and Changbom~Park\inst{4}
\and Mirt~Gramann\inst{1} 
\and Pekka~Hein\"am\"aki\inst{5}
\and Enn~Saar\inst{1,6}
\and Jaan~Einasto\inst{1,6,7}
}
\institute{Tartu Observatory, Observatooriumi 1, 61602 T\~oravere, Estonia
\and
Institute of Physics, University of Tartu, W. Ostwaldi 1, 50411, Tartu, Estonia
\and
Leibniz-Institut f\"ur Astrophysik Potsdam (AIP), An der Sternwarte 16, D-14482
Potsdam, Germany
\and 
School of Physics, Korea Institute for Advanced Study, 85 Hoegiro, Dong-Dae-Mun-Gu, Seoul 02455, Korea
\and
Tuorla Observatory, University of Turku, V\"ais\"al\"antie 20, Piikki\"o, Finland
\and 
Estonian Academy of Sciences, Kohtu 6, 10130 Tallinn, Estonia
\and
ICRANet, Piazza della Repubblica 10, 65122 Pescara, Italy
}

\authorrunning{Einasto, M. et al. }

\offprints{Einasto, M.}

\date{ Received   / Accepted   }

\titlerunning{A2142}

\abstract
{
Superclusters of galaxies provide dynamical environments for the study of the
formation and evolution of structures in the cosmic web from galaxies, to the
richest galaxy clusters, and superclusters themselves.
}
{
We study galaxy populations and search for possible merging substructures in the rich galaxy 
cluster \object{A2142} in the collapsing core of the supercluster 
SCl~A2142, which may give rise to radio and X-ray structures in the cluster, 
and affect galaxy properties of this cluster.
}
{
We used normal mixture modelling to select substructure of the cluster A2142. We compared
alignments of the cluster, its brightest galaxies (hereafter BCGs), 
subclusters, and supercluster axes. 
The projected phase space (PPS) diagram and clustercentric distributions are used
to analyse the dynamics of the cluster and study the 
distribution of various galaxy populations in the cluster and subclusters. 
}
{We find several infalling galaxy groups and subclusters. 
The cluster, supercluster, BCGs, and one infalling subcluster are all aligned.
Their orientation is correlated with the alignment of the radio and X-ray haloes
of the cluster. 
Galaxy populations in the main cluster and in the outskirts subclusters are different. 
Galaxies in the centre of the main cluster  at the clustercentric distances
$0.5$~\Mpc\ 
($D_{\mathrm{c}}/R_{\mathrm{vir}} < 0.5$, $R_{\mathrm{vir}} = 0.9$~\Mpc) 
have older stellar
populations (with the median age of $10 - 11$~Gyrs) than galaxies
at larger clustercentric distances. 
Star-forming and recently quenched galaxies are located mostly
at the clustercentric distances $D_{\mathrm{c}} \approx 1.8$~\Mpc,\ 
where subclusters fall into the cluster and  
the properties of 
galaxies change rapidly. In this region the median age of stellar populations of 
galaxies is about $2$~Gyrs.   
Galaxies in A2142 on average have higher stellar masses, lower star formation rates, 
and redder colours than galaxies in rich groups.
The total mass in infalling groups and subclusters is 
$M \approx 6\times10^{14}h^{-1}M_\odot$, that is approximately half of the mass
of the cluster. This mass is sufficient
for the mass growth of the cluster  
from redshift $z = 0.5$ (half-mass epoch) to the present.
}
{Our analysis suggests that the cluster A2142 has formed as a 
result of past and present mergers and infallen groups, predominantly along the
supercluster axis. Mergers cause complex
radio and X-ray structure of the cluster and affect the properties of  galaxies
in the cluster, especially at the boundaries of the cluster in the  
infall region.  
Explaining the differences between galaxy populations, mass, and richness
of  A2142, and other groups and clusters
may lead to better insight about
the formation and evolution of rich galaxy clusters.
}


\maketitle

\section{Introduction} 
\label{sect:intro} 

The cosmic web, evolved from tiny 
density perturbations in the very early Universe,
consists of galaxies and galaxy systems from the smallest groups 
to the richest superclusters, separated by voids
\citep{1978MNRAS.185..357J, 1988Natur.334..129K}. 

Early studies
of the cosmic web have already shown that galaxy superclusters have important 
morphological property: the most luminous galaxy clusters 
are typically located in the high-density core
regions of rich superclusters at the crossing of 
several cluster or galaxy filament chains  \citep{Joeveer:1978a,1978MNRAS.185..357J}.
These are locations in which rich galaxy clusters form via merging and accretion 
of smaller structures (galaxies and groups of galaxies) along filaments 
\citep[][and references therein]{1996Natur.380..603B, 2009LNP...665..291V,
2011A&A...531A.149S, 2012ARA&A..50..353K}.

The high-density cores of rich superclusters 
are the largest objects in the Universe that may collapse now or during future evolution \citep{1998ApJ...492...45S, 
2000AJ....120..523R, 2006A&A...447..133P,
2014MNRAS.441.1601P, 2015A&A...581A.135G, 2015MNRAS.453..868O, 2015A&A...575L..14C, 
2016A&A...595A..70E}. These cores form an evolving environment
for the study of the properties of galaxies, groups, and clusters inside
them.
It is especially interesting to investigate rich galaxy clusters 
in the collapsing
cores of superclusters, where we can study various processes
and transformation of galaxies both in clusters and their outskirts, and in their
larger scale environment. These processes are responsible for transformation
of galaxies from mostly blue, star-forming field galaxies to red, quiescent 
cluster population \citep[see][for a review and references]
{2015ApJ...806..101H, 2017ApJ...843..128R}.

Merging clusters in the collapsing cores of rich superclusters have been
studied, for example, in the Shapley supercluster \citep{2015MNRAS.446..803M}
and in the A2199 supercluster \citep{2017ApJ...842...88S}.
One of such clusters is the very rich galaxy cluster A2142 embedded in the
collapsing core of the supercluster SCl~A2142
\citep{2015A&A...580A..69E, 2015A&A...581A.135G}.

The structure of the central region of the cluster A2142 up to virial radius 
has been studied in detail
in optical \citep{2011ApJ...741..122O, 
2014A&A...566A..68M}, 
radio
\citep{2010A&A...522A.105G, 2013ApJ...779..189F, 2017A&A...603A.125V} and X-ray (also joint analysis
with optical bands)
wavelengths \citep{2000ApJ...541..542M, 2001cghr.confE..33F, 2008PASJ...60..345O,
2013A&A...556A..44R, 2014A&A...570A.119E, 2016A&A...595A..42T,
2017A&A...605A..25E}. These studies revealed that this cluster is a
cold front cluster 
with signs of gas sloshing and giant multiple radio haloes that have been
interpreted as signs of multiple mergers. \citet{2014A&A...570A.119E,
2017A&A...605A..25E} discovered a stripped galaxy group infalling into 
A2142.  \citet{2013A&A...556A..44R} showed that in A2142 (merger induced) gas sloshing
occurs at scales up to about $1$~\Mpc, which is much larger than typically
in clusters and larger than obtained from simulations. 
A recent analysis of radio data of A2142 suggest several mergers and/or
infalling galaxy groups in A2142 \citep{2017A&A...603A.125V}.
The  cluster A2142 is unusually rich; it is, for example,  
twice as rich as the richest galaxy clusters in the Sloan Great Wall
(SGW) approximately at the same distance \citep{2010A&A...522A..92E, 2014A&A...566A...1T}.
\citet{2016ApJ...827L...5M} 
showed that 
in simulations massive haloes have
significantly lower amount of massive subhaloes in comparison with  A2142. 

In this paper we analyse the substructure and galaxy content 
of the cluster A2142 and its outskirts at clustercentric distances 
up to about $3$~\Mpc. 
We present an analysis of the full supercluster SCl~A2142 in 
a forthcoming study. 
Our aim is to search for possible substructures that may give rise
of the radio and X-ray structures of A2142 cluster and affect galaxy properties in the cluster.
We determine substructures in the cluster and its outskirts
and analyse their orientations. We study galaxy populations of the cluster and
subclusters in the projected phase space 
 (PPS) diagram  and as a function of clustercentric distance.
Recently, the analysis of the PPS diagram has become an important tool to
study galaxy populations in galaxy groups, clusters, and multiclusters
\citep{2015ApJ...806..101H, 2015MNRAS.448.1715J,
2017MNRAS.467.4410A, 2017ApJ...838..148P, 2017ApJ...838...81Y, 2017ApJ...843..128R,
2017MNRAS.471..182W}. In the PPS diagram galaxies with 
different accretion histories populate different areas. This can be used
to analyse dynamical properties of galaxies in clusters and outskirts and
to compare galaxy populations in virialised and nonvirialised regions
of the clusters.
We also compare galaxy  populations in A2142 and in other 
rich galaxy groups and clusters from the Sloan Digital Sky Survey.
 
We assume  the standard cosmological parameters: the Hubble parameter $H_0=100~ 
h$ km~s$^{-1}$ Mpc$^{-1}$, matter density $\Omega_{\rm m} = 0.27$, and 
dark energy density $\Omega_{\Lambda} = 0.73$ \citep{2011ApJS..192...18K}.

\section{Data} 
\label{sect:data} 

\subsection{Supercluster, group, and filament data}
\label{sect:gr}

Our initial dataset is selected from 
the MAIN sample of the 10th data release of the Sloan Digital Sky 
Survey (SDSS) \citep{2011ApJS..193...29A, 2014ApJS..211...17A}.
We used a spectroscopic galaxy sample  with the 
apparent Galactic extinction corrected $r$ magnitudes $r \leq 
17.77$ and redshifts $0.009 \leq z \leq 0.200$. 
We corrected the redshifts of galaxies for the motion relative to the cosmic microwave background and 
computed the comoving distances of galaxies \citep{2002sgd..book.....M}. 
Galaxies with unreliable parameters were removed
from the sample as described in detail 
in \citet{2012A&A...540A.106T} and \citet{2014A&A...566A...1T}.
The SDSS spectroscopic sample is incomplete because of fibre collisions; the 
smallest separation between spectroscopic fibres is 55", and approximately
6\% of the potential targets for spectroscopy are without observed spectra because of this.
Below we analyse the completeness of the  data 
used in this paper  due to fibre collisions.

The SDSS MAIN dataset was used to calculate the luminosity-density
field and detect superclusters of galaxies in this field
to find groups of galaxies with the friends-of-friends algorithm
and to determine galaxy filaments by applying the Bisous process to the
distribution of galaxies \citep{2014MNRAS.438.3465T}. 
The data from supercluster, group, and filament catalogues were then
used to select galaxy group and filament information for the 
supercluster SCl~A2142.

{\it Supercluster catalogue and supercluster SCl~A2142.}
Data on SCl~A2142 are taken from the catalogue
of galaxy superclusters in which superclusters were defined as
extended connected volumes above a threshold density level
$5.0$
(in units of mean density, $\ell_{\mathrm{mean}}$ = 
1.65$\cdot10^{-2}$ $\frac{10^{10} h^{-2} L_\odot}{(\vmh)^3}$)
in the luminosity density field  \citep{2012A&A...539A..80L}.

The supercluster SCl~A2142 \citep[SCl~001 in ][]{2012A&A...539A..80L}
at redshift $z \approx 0.09$ has over a thousand member galaxies; the total length of this supercluster, defined as the maximum distance between galaxy pairs
in the supercluster, is $\approx 50$~\Mpc\ \citep{2012A&A...539A..80L}.
This supercluster was recently described 
in \citet{2015A&A...580A..69E} and \citet{2015A&A...581A.135G} 
who showed that the high-density core
of this supercluster, with a radius $\approx 13$~\Mpc\
and mass of about $M \approx 4\times~10^{15}h^{-1}M_\odot$, is already collapsing.

\begin{figure}[ht]
\centering
\resizebox{0.44\textwidth}{!}{\includegraphics[angle=0]{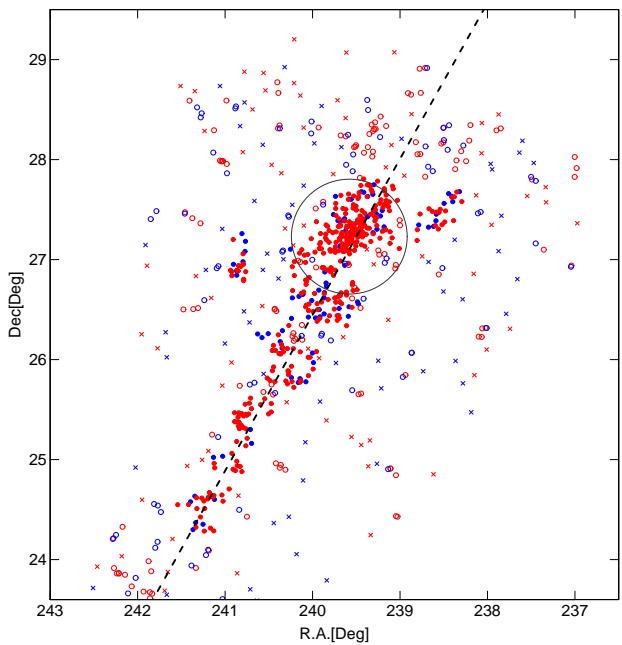}}
\caption{
Sky distribution of galaxies in SCl~A2142. 
The red symbols show galaxies with old stellar populations
($D_n(4000) \geq 1.55$), and the blue symbols denote galaxies with young 
stellar populations ($D_n(4000) < 1.55$).
The filled circles correspond to galaxies in rich groups with at least 10 member galaxies,
the empty circles indicate galaxies in poor groups with $2 - 9$ galaxies,
and the crosses denote single galaxies. The grey circle has a radius of about $3$~\Mpc\
and denotes the region of the cluster A2142 and its outskirts.
The dashed line shows the supercluster axis.
}
\label{fig:radec}
\end{figure}

\begin{figure}[ht]
\centering
\resizebox{0.44\textwidth}{!}{\includegraphics[angle=0]{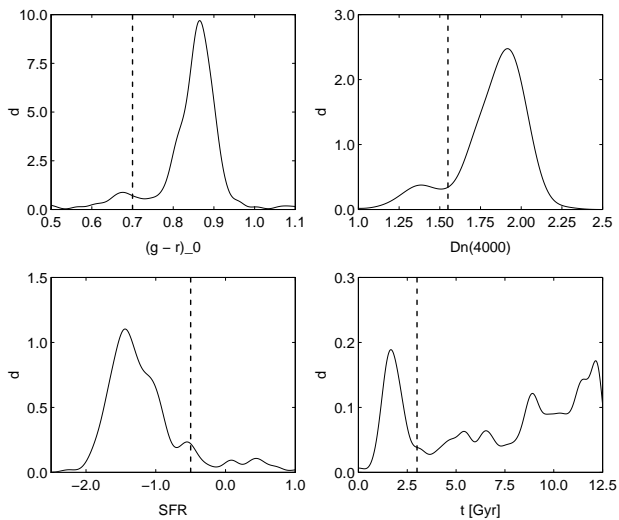}}
\caption{
Distribution of galaxy colours  $(g - r)_0$ (upper left panel),
$D_n(4000)$ index (upper right panel), star formation rates  $\log \mathrm{SFR}$
(lower left panel), and stellar ages $t$ (lower right panel) 
in the cluster A2142.
The dashed lines show the parameter values used to divide galaxies into
populations as described in the text.
}
\label{fig:a21424para}
\end{figure}

\begin{figure}[ht]
\centering
\resizebox{0.44\textwidth}{!}{\includegraphics[angle=0]{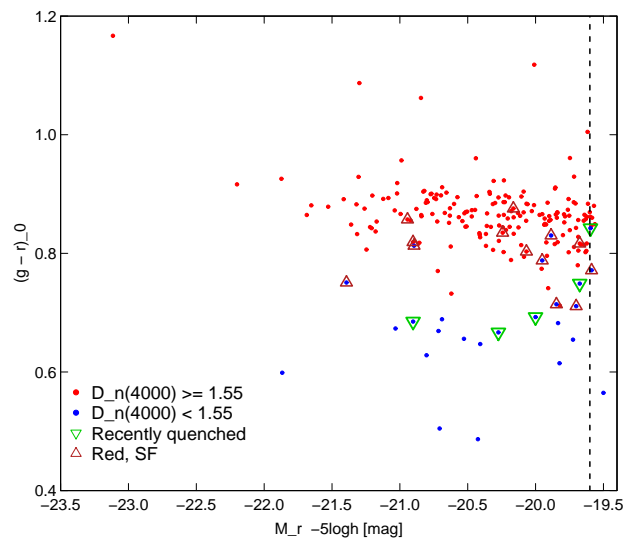}}
\caption{
Colour - magnitude diagram $(g - r)_0$ vs. $M_r$ for galaxies in the cluster A2142.
The red filled circles correspond to the galaxies with old stellar populations
($D_n(4000) \geq 1.55$) and the blue filled circles to the galaxies having young 
stellar populations with $D_n(4000) < 1.55$. The green triangles show 
recently quenched  galaxies 
with $D_n(4000) \leq 1.55$ and star formation rate $\log \mathrm{SFR} < -0.5$.
The red triangles indicate red, high SFR galaxies  defined as 
$g - r \geq 0.7$, and $\log \mathrm{SFR} \geq -0.5$.
The dashed line shows the completeness limit $M_r = -19.6$~$+5\log_{10}h$.
}
\label{fig:a2142cmr}
\end{figure}

{\it Galaxy groups  in the supercluster SCl~A2142.}
We selected galaxy groups belonging to SCl~A2142 
from the group catalogue by \citet{2014A&A...566A...1T}.
In this catalogue galaxy groups were  determined using 
the friends-of-friends cluster analysis 
method introduced in cosmology by \citet{1982Natur.300..407Z} and 
\citet{1982ApJ...257..423H}. A galaxy belongs to a 
group of galaxies if this galaxy has at least one group member galaxy closer 
than a linking length. In a flux-limited sample the density of galaxies slowly 
decreases with distance. To take this selection effect into account properly 
when constructing a group catalogue from a flux-limited sample, the 
linking length was rescaled with distance, calibrating the scaling relation by observed 
groups. As a result, the 
maximum sizes in the sky projection and the velocity dispersions of 
\citet{2014A&A...566A...1T} groups 
are similar at all distances. 
The details about the data 
reduction, group finding procedure, and description of the group catalogue can be found in 
\citet{2014A&A...566A...1T}.

SCl~A2142 embeds 14 galaxy groups with at least 10 member
galaxies. In the present study, we 
only used the data of the cluster A2142 
(Gr3070 in the catalogue) (Table~\ref{tab:gr10}). 
The full supercluster SCl~A2142  covers only a narrow distance interval
($\approx 255 - 275$~\Mpc), and therefore we used data about the group from flux-limited sample 
\citep[see also ][for details of data selection]{2015A&A...580A..69E}.
In Table~\ref{tab:gr10} we also give data on A2142 components
and subclusters, defined below in Sect.~\ref{sect:str}.
Masses  
of the cluster A2142 and its components in Table~\ref{tab:gr10}
were  calculated as described
in \citet{2014A&A...566A...1T}. These authors applied the virial theorem, 
assuming symmetry of galaxy velocity distribution 
and the Navarro-Frenk-White (NFW)
density profile for galaxy distribution in the plane of the sky. 
For a detailed description of how the masses  were calculated,
we refer to \citet{2014A&A...566A...1T}.
The virial radius of the cluster, $R_\mathrm{vir} = 0.9$~\Mpc.
It is calculated as
\begin{equation}\label{eq:rvir}
        \frac{1}{R_\mathrm{vir}} = \frac{2}{(1+z_\mathrm{m})n(n-1)}\sum\limits^{n}_{i\neq j}\frac{1}{R_{ij}},
\end{equation}
where $R_{ij}$ is the projected distance between galaxies in pairs in a group,
$z_\mathrm{m}$ is the mean redshift of the group, and $n$ is the number of 
galaxies in a group \citep[see][for details]{2014A&A...566A...1T}.
Figure~\ref{fig:radec} shows
the sky distribution of galaxies in SCl~A2142. 
We plot galaxies from rich and poor groups with 
filled and empty circles, correspondingly. 
Sky coordinates of galaxies from rich groups with at least 10 member
galaxies have been used below to determine the orientation of the supercluster
axis.  We denote the region of the cluster A2142 and its outskirts
analysed in this study.

Absolute magnitudes of galaxies are computed according to the formula
\begin{equation}
M_r = m_r - 25 -5\log_{10}(d_L)-K,
\end{equation} 
where $d_L$ is the luminosity distance in units of $h^{-1}$Mpc and
$K$ is the $k$+$e$-correction. 
The $k$-corrections were calculated with the \mbox{KCORRECT\,(v4\_2)} code
\citep{2007AJ....133..734B} and the evolution corrections have been calibrated 
according to \citet{2003ApJ...592..819B}. 
Details about how the $k$ and $e$-corrections were applied 
can be found in \citet{2014A&A...566A...1T}.
\citet{2014A&A...566A...1T} used  slightly smaller
evolution corrections than  \citet{2003ApJ...592..819B}.
The value of  $M_{\odot} = 4.53$ (in $r$-filter).

At the distance of A2142, $ \approx 265$~\Mpc, the sample is complete
at the absolute magnitude limit $M_r = -19.6$ in units of $\mathrm{mag}+5\log_{10}h$. 
In our sample there is six
galaxies fainter than this limit. In our study we used the full dataset
of A2142 to determine the substructure in A2142, and for statistical analysis
we used a magnitude-limited complete sample, excluding galaxies fainter than
the completeness limit, $M_r = -19.6$~$+5\log_{10}h$.  
 In addition, we checked the data
 in the region covered by the cluster A2412 (Gr3070) used in this paper 
 and found that in this region no galaxy is missing because of the fibre collision effect.

\begin{table}[ht]
\caption{Data for the Abell cluster A2142 (Gr3070) components.}
\begin{tabular}{rrrrrrr} 
\hline\hline  
(1)&(2)&(3)&(4)&(5)& (6)&(7)\\      
\hline 
 $ID$ & $N_{\mathrm{gal}}$&  $\mathrm{R.A.}$ & $\mathrm{Dec.}$  & $\mathrm{Dist.}$ 
 &  $M$ & $D_n(4000)$\\
 && [deg] & [deg] & [$Mpc~h^{-1}]$ &   &\\
\hline
$A2142$ & 212 & 239.5 & 27.3 & 265 & 9.1 &  1.9\\
 $C1$   & 135 & 239.6 & 27.2 & 266 & 7.1 & 1.9\\
 $C2$   &  64 & 239.3 & 27.5 & 262 & 2.0 &  \\
 \hline                                      
 $M1$   &  22 & 239.3 & 27.3 & 257 &  .9 & 1.9 \\
 $M2$   &  17 & 239.4 & 27.7 & 262 &  .9 & 1.7 \\
 $M3$   &  24 & 239.2 & 27.6 & 262 & 1.5  & 1.8\\
 \hline                                       
 $C3$   &   8 & 239.8 & 26.9 & 255 &  1.5 & 1.9\\
 \hline
\label{tab:gr10}  
\end{tabular}\\
\tablefoot{                                                                                 
Columns are as follows:
(1): Component ID \citep[][]{2015A&A...580A..69E};
(2): the number of galaxies in a component;
(3--5) R.A., Dec., and distance of a component centre; 
(6): the mass of a component, in units of $10^{14}h^{-1}M_\odot$; and
(7): median value of the $D_n(4000)$ index (see Sect.~\ref{sect:gal}).
}
\end{table}

\begin{figure*}[ht]
\centering
\resizebox{0.58\textwidth}{!}{\includegraphics[angle=0]{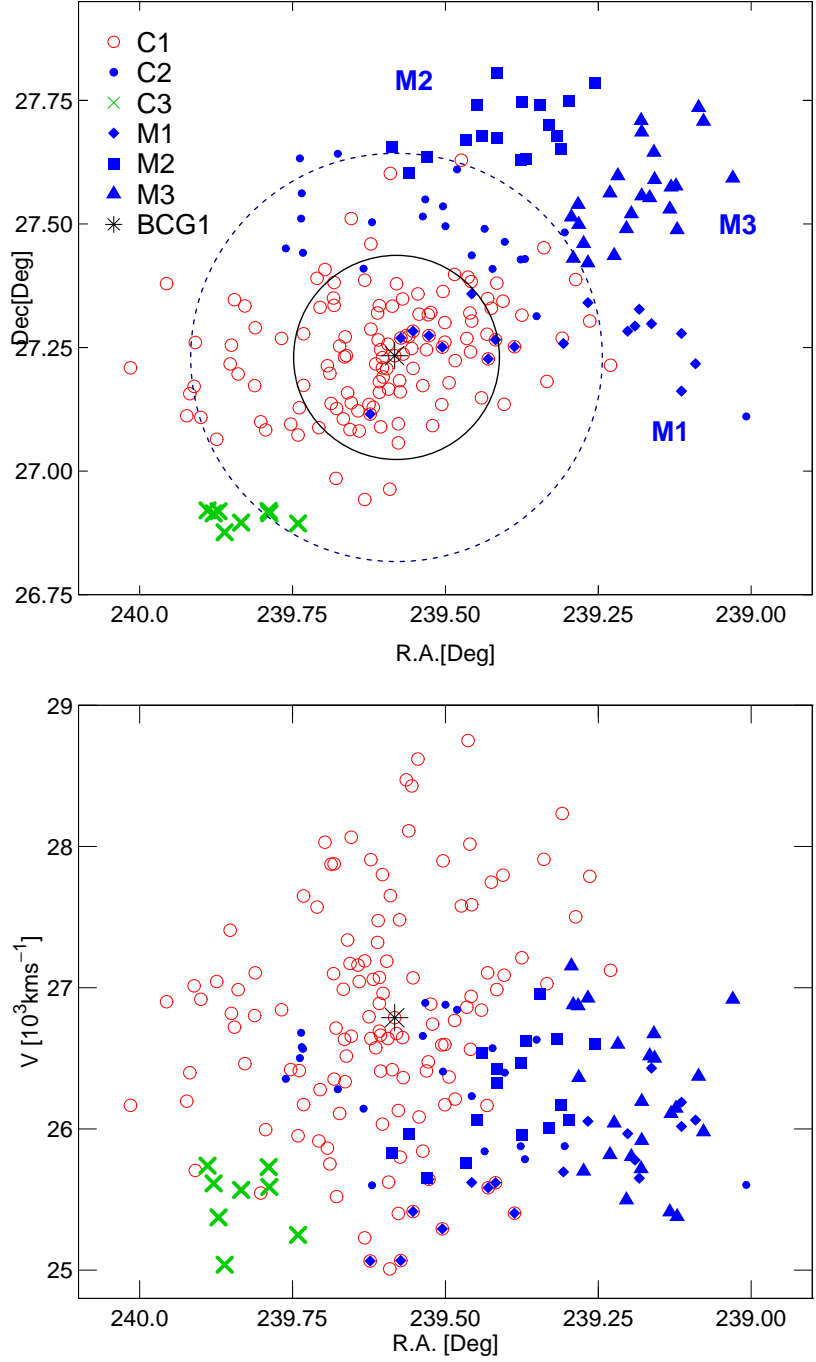}}
\caption{
Distribution of galaxies  in the cluster A2142 and its outskirts
in the sky plane (upper panel)
and in the R.A. - velocity plane (lower panel)
in three components found by {\it mclust}. 
Red symbols correspond to the galaxies in the first component, C1.
Blue symbols show galaxies from the second component, C2, and
green symbols denote the location of galaxies from the third
component, C3. 
Galaxies from the component C2 are  divided into subclusters M1 (diamonds),
M2 (squares), M3 (triangles), and dots (those component C2 galaxies
that are not associated with subclusters) as explained in the text.
C1 member galaxies  assigned  to the subcluster M1
are shown with red circles with blue diamonds (see text). 
The radius of the small black circle 
corresponds to the virial radius of the cluster, $R_{\mathrm{vir}} = 0.9$~\Mpc,
and the radius of the large blue circle (dotted line) corresponds
to the radius of the component C1 (main cluster), $R_{\mathrm{max}} = 1.8$~\Mpc.
}
\label{fig:a2142mclust}
\end{figure*}

\begin{figure*}[ht]
\centering
\resizebox{0.55\textwidth}{!}{\includegraphics[angle=0]{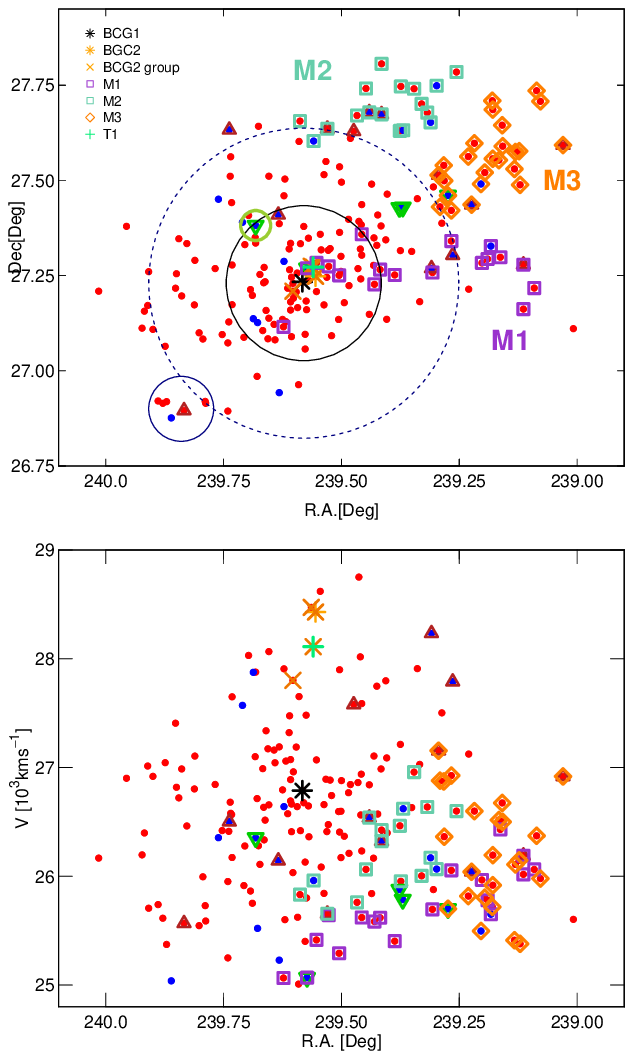}}
\caption{
Distribution of galaxies  in the cluster A2142 and its outskirts
in the sky plane (upper panel)
and in the R.A. - velocity plane (lower panel). 
Red filled circles correspond to the galaxies with old stellar populations
($D_n(4000) \geq 1.55$) and blue filled circles to the galaxies that have young 
stellar populations with $D_n(4000) < 1.55$. Green triangles show 
recently quenched  galaxies 
with $D_n(4000) \leq 1.55$ and star formation rate $\log \mathrm{SFR} < -0.5$.
Red triangles indicate red, high SFR galaxies  defined as 
$g - r \geq 0.7$, and $\log \mathrm{SFR} \geq -0.5$
(as in Fig.~\ref{fig:a2142cmr}).
The black star shows the location of the brightest cluster galaxy (BCG1),
and the orange star shows the location of the second BCG (BCG2). The orange crosses
show galaxies from a possible BCG2 group. The green cross near the centre
denotes a galaxy with a radio tail \citep[T1 in ][]{2017A&A...603A.125V}. 
The dark violet squares indicate galaxies from subcluster M1,
the aquamarine squares indicate galaxies from subcluster M2, and the orange diamonds
show galaxies belonging to the third subcluster, M3.  
The small green circle shows the location of galaxies from infalling group $G_E$
described in \citet{2014A&A...570A.119E},
and the navy circle shows the location of galaxies from the poor group, which
corresponds to the third component of the cluster \citep{2015A&A...580A..69E}.  
The radius of the black circle 
corresponds to the virial radius of the cluster, $R_{\mathrm{vir}} = 0.9$~\Mpc,
and the radius of the large blue circle (dotted line) corresponds
to the radius of the component C1 (main cluster), $R_{\mathrm{max}} = 1.8$~\Mpc.
}
\label{fig:a2142radec}
\end{figure*}

{\it Galaxy filaments.}
Galaxy filaments belonging to the supercluster SCl~A2142
were extracted from the catalogue of galaxy filaments  by \citet{2014MNRAS.438.3465T}.
This catalogue was built by applying the Bisous process
to the distribution of galaxies \citep{2016A&C....16...17T}.
Details about how the galaxy filaments in SCl~A2142 were selected 
were described in \citet{2015A&A...580A..69E}. 
In the present study, we only used the data about filaments
in the neighbourhood of the cluster A2142 to check for the filament
membership of galaxies. Combining data from different datasets
(supercluster, group, and filament catalogues) gives us
more comprehensive information about the cluster and its outskirts.
Catalogues of galaxy superclusters, groups, and filaments are
available from the database of cosmology-related catalogues 
at \url{http://cosmodb.to.ee/}.

\subsection{Galaxy populations}
\label{subsec:galpop}

Data about galaxies were downloaded from the SDSS DR10 
web page\footnote{\url{http://skyserver.sdss3.org/dr10/en/help/browser/browser.aspx}. }
Galaxy magnitudes and  Petrosian radii were taken from the SDSS spectroscopic and 
photometric data, correspondingly.
We calculated galaxy colours as $(g - r)_0 = M_g - M_r$. 
All magnitudes and colours correspond to  the rest frame at  redshift $z=0$.
The concentration index of galaxies is calculated as the ratio of the 
Petrosian radii 
$R_{50}$ and  $R_{90}$: $R_{50}/R_{90}$.

Galaxy stellar masses, star formation rates (SFRs), and 
$D_n(4000)$ index of galaxies  
are from the MPA-JHU spectroscopic catalogue \citep{2004ApJ...613..898T, 
2004MNRAS.351.1151B}, from which the various properties of 
galaxies were obtained by fitting SDSS photometry and spectra with
the stellar population synthesis models developed by \citet{2003MNRAS.344.1000B}.
The stellar masses of galaxies were derived as 
described by \citet{2003MNRAS.341...33K}.  The SFRs 
were computed using the photometry and emission lines as described 
by \citet{2004MNRAS.351.1151B} and \citet{2007ApJS..173..267S}. 
The strength of the $D_n(4000)$ break (the ratio of the average flux densities
in the band $4000 - 4100 \angstrom$ and $3850 - 3950 \angstrom$; $D_n(4000)$ index)
is correlated with the time passed from the most recent star formation event 
and is defined as in \citet{1999ApJ...527...54B}.\ The $D_n(4000)$ index characterises star formation histories of galaxies.

The stellar velocity dispersions
were obtained by fitting galaxies with the publicly available Gas AND Absorption Line Fitting 
\citep[GANDALF; ][]{2006MNRAS.366.1151S}
and penalised PiXel Fitting \citep[pPXF; ][]{2004PASP..116..138C}. 
Stellar ages are from the Portsmouth group \citep{2009MNRAS.394L.107M}.

In Fig.~\ref{fig:a21424para} we plot the distributions of galaxy colours  $(g - r)_0$,
$D_n(4000)$ index, star formation rates  $\log \mathrm{SFR}$,
and stellar ages $t$ in the cluster A2142. 

The value $(g - r)_0 = 0.7$ is used to separate red and blue galaxies, where red galaxies 
have $(g - r)_0 \geq 0.7$. 
The percentage of red galaxies in A2142 is $f_{\mathrm{red}} = 0.92$.

The $D_n(4000)$ index was used to separate quiescent and star-forming
galaxies. We applied values for  quiescent galaxies with old stellar populations 
as having  $D_n(4000) \geq 1.55$ (89\% of all galaxies from the complete sample). 
Star-forming galaxies with young stellar 
populations have $D_n(4000) < 1.55$. This  
limit was also used by \citet{2003MNRAS.341...54K} and \citet{2017A&A...605A...4H} 
to separate young and old galaxies 
in the SDSS survey. 

We divided galaxies with old and young stellar populations by stellar age
using the age limit $t = 3$~Gyrs.
Galaxies with $t \geq 3$~Gyrs formed 77\%
of all galaxies. The SFR  $\log \mathrm{SFR} \leq -0.5$ 
corresponds to quiescent galaxies (88\% of galaxies). 
Actively star-forming galaxies are characterised by 
$\log \mathrm{SFR} > -0.5$; this limit was also applied in   
\citet{2014A&A...562A..87E}.

We might use slightly higher values of the colour index $(g - r)_0$
and star formation rate $\log \mathrm{SFR}$ to separate 
galaxy populations, this only changes the percentages of red
and star-forming galaxies marginally.
For example, if for red galaxies $(g - r)_0 \geq 0.75$
then the percentage of red galaxies in A2142 becomes  $f_{\mathrm{red}} = 0.89$.
Present values were used consistently with earlier studies
\citep[see, for example, ][]{2014A&A...562A..87E, 2017A&A...605A...4H}.

Some star-forming galaxies have red colours and are known as red star-forming
galaxies. Also, there are galaxies with low SFRs
but also
low values of $D_n(4000)$ index, which suggest that they may be recently quenched.
In our sample  
we define recently quenched galaxies as those
with $D_n(4000) \leq 1.55$ and  $\log \mathrm{SFR} < -0.5$ (five galaxies).
Red, high SFR galaxies are defined as galaxies with 
$g - r \geq 0.7$ and $\log \mathrm{SFR} \geq -0.5$ (13 galaxies). 
 
The A2142 colour - magnitude diagram is shown in Fig.~\ref{fig:a2142cmr}.
We plot galaxies with old and young stellar populations, recently quenched
galaxies, and red, star-forming galaxies with different symbols.
Here we show also the completeness limit in luminosity,
$M_r = -19.6$~$+5\log_{10}h$. As seen also from this figure, most galaxies in A2142 are red,
including two  recently quenched galaxies.
Among galaxies that are fainter than the completeness limit only one is blue.

\section{Analysis}
\label{sect:results} 

\subsection{Structure of the cluster A2142 and its outskirts}
\label{sect:str} 
The structure of the cluster A2142 has been analysed in
\citet{2012A&A...540A.123E} with a number of
1D, 2D, and 3D tests. Both 3D and 2D  analyses revealed that A2142
has three components. The distribution of galaxy velocities
in the components overlap, so that 1D methods do not find significant
substructures in the cluster. The details of the
tests can be found in \citet{2012A&A...540A.123E}.
In the present study we focus on the 3D analysis of the cluster structure,
following \citet{2015A&A...580A..69E} who
used multidimensional normal mixture modelling 
to separate the main A2142 cluster and its outskirts systems.

To search for components in A2142, {\it mclust} 
package for classification and clustering
\citep{fraley2006} from {\it R}, an open-source free statistical environment 
developed under the GNU GPL \citep[][\texttt{http://www.r-project.org}]{ig96}
was applied. This package is based on the analysis of a finite mixture of 
distributions, 
in which each mixture component is taken to correspond to a 
different subgroup of the cluster.
As an input for {\it mclust} we used 
the sky coordinates and velocity of the cluster A2142 member galaxies. 
The values of velocities were scaled to make them of the same order as the values
of coordinates.  The best solution for components found by {\it mclust}
was chosen using the Bayesian information criterion (BIC). 
To test how the possible errors in the line-of-sight positions
of galaxies affect the results of component analysis, the velocities
of galaxies were shifted randomly 1000 times and each time  {\it mclust}
was run to search for components. The size of the shifts was chosen from a Gaussian
distribution with the dispersion equal to the velocity distribution
of galaxies in the cluster. The number of components found by
{\it mclust} remained unchanged showing that the component finding 
was not sensitive to such errors.  
For details of the analysis of the cluster A2142 with {\it mclust},
we refer to \citet{2012A&A...540A.123E} and \citet{2015A&A...580A..69E}.

We  show the results of the analysis of the substructure in A2142  with {\it mclust}
in Fig.~\ref{fig:a2142mclust}, where we plot
the distribution of galaxies in the cluster A2142 in the plane 
of the sky and in the sky-velocity plane, and show different components
in A2142 and its outskirts region. 
Velocities of galaxies have been calculated from their comoving distances.

In A2142 {\it mclust} revealed three components, C1 - C3 (Table~\ref{tab:gr10}).
The distribution of galaxy velocities
in the components largely overlap (Fig.~\ref{fig:a2142mclust}).
The richest of the components, C1, corresponds  to the main cluster
with the maximal radius in the sky, $R_{\mathrm{max}} = 1.8$~\Mpc. 

Galaxies from the second component, C2, plotted in blue in Fig.~\ref{fig:a2142mclust},
have velocity distributions similar to the galaxies from the main cluster (C1).
As seen in the upper panel of Fig.~\ref{fig:a2142mclust}, in the sky plane
they are located in the outskirts of the main cluster.  
To use as much information as possible about the structures in the outskirts region
of the cluster,
our next step was to compare the components found with {\it mclust}
with the data from filament catalogue by \citet{2014MNRAS.438.3465T}
that was briefly described in Sect.~\ref{sect:gr}. This comparison showed that
close to the main cluster (our component C1) the filament catalogue
lists two short elongated structures. Galaxies from these structures 
are mostly members
of the second component, C2. These galaxies are denoted as subclusters M1 and M2 in 
Fig.~\ref{fig:a2142mclust}.
The filaments
very close to the clusters are not well defined, 
and therefore we refer to these systems as "subclusters". 
Some galaxies from the first subcluster, M1, are projected inside the
virial radius of the main cluster and, according to {\it mclust},
belong to the component C1 (see Fig.~\ref{fig:a2142mclust}), 
but following the previous analysis
we assign these galaxies to the subcluster M1. 
Figure~\ref{fig:a2142mclust} shows that in the outskirts of the cluster
there are galaxies between subclusters M1 and M2. In what follows, we 
assign these galaxies to the third subcluster, M3. 
Below we show that subclusters M2 and M3 have different galaxy content
and orientations than M1, 
which suggests that they are different subclusters.

In the outskirts of the cluster {\it mclust} also found a third, very poor component, C3.
This component has only eight member galaxies. 

Summarising, Fig.~\ref{fig:a2142mclust} shows that the structure of the cluster and its
outskirts is rather complex, having several subclusters, revealed using 
combined results of different structure finding algorithms. 
We continue the analysis of these structures and galaxy populations in them below.

In addition to this analysis, previous studies have identified 
 several possible infalling groups in A2142. We show them in
Fig.~\ref{fig:a2142radec} where we plot
the distribution of galaxies in the cluster A2142 and its subclusters 
in the plane 
of the sky and in the sky-velocity plane and use different colours for 
galaxies with old and young stellar populations.

In Fig.~\ref{fig:a2142radec} the location of the two brightest galaxies 
in the cluster (BCG1 and BCG2) is denoted with stars. 
The lower panel of Fig.~\ref{fig:a2142radec} 
shows that the BCG2 has high 
velocity with respect to the mean velocity of the cluster
\citep[see also][]{2000ApJ...541..542M, 2011ApJ...741..122O}. This galaxy may be the brightest
galaxy of merged group or cluster 
\citep[][]{2000ApJ...541..542M}. Other possible 
members of the BCG2 group that have very close sky positions and  
similar velocities as the BCG2 are also indicated. These members 
correspond to the substructure S1 in \citet{2011ApJ...741..122O}.  
Moreover, \citet{2017A&A...603A.125V} have described two long tailed radio sources, 
named T1 and T2 \citep[see also][]{2010A&A...522A.105G}. 
The T1 galaxy is one of the possible BGC2 group members that is mentioned  in
\citet{2000ApJ...541..542M}.
We indicate this galaxy in Fig.~\ref{fig:a2142radec} as well. 

\citet{2014A&A...570A.119E, 2017A&A...605A..25E}
detected a galaxy group infalling into A2142.
Some galaxies from our sample lie at the edge of this group,
denoted as  $G_E$.
We indicate
the location of these galaxies in Fig.~\ref{fig:a2142radec}
with a green circle.

\begin{figure}[ht]
\centering
\resizebox{0.44\textwidth}{!}{\includegraphics[angle=0]{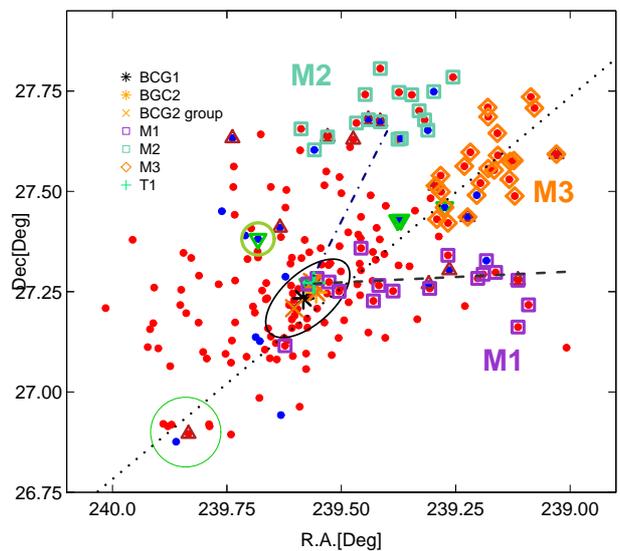}}
\caption{
Distribution of galaxies  in the cluster A2142 and its outskirts
in the sky plane. The symbols are the same as in Fig.~\ref{fig:a2142radec}.
The lines show the direction of supercluster axis (dotted line),
subcluster M1 axis (dashed line,
this direction coincides with the direction of radio tail of
the galaxy T1, see text), and subcluster M2 direction (dash-dotted line,
coincides with the direction of radio tail of
the galaxy T2). 
The ellipse shows approximately the contour of the X-ray halo
of the cluster and radio emission at 1.38 GHz 
\citep[see][for details]{2017A&A...603A.125V}.
}
\label{fig:axes}
\end{figure}

\begin{figure}[ht]
\centering
\resizebox{0.46\textwidth}{!}{\includegraphics[angle=0]{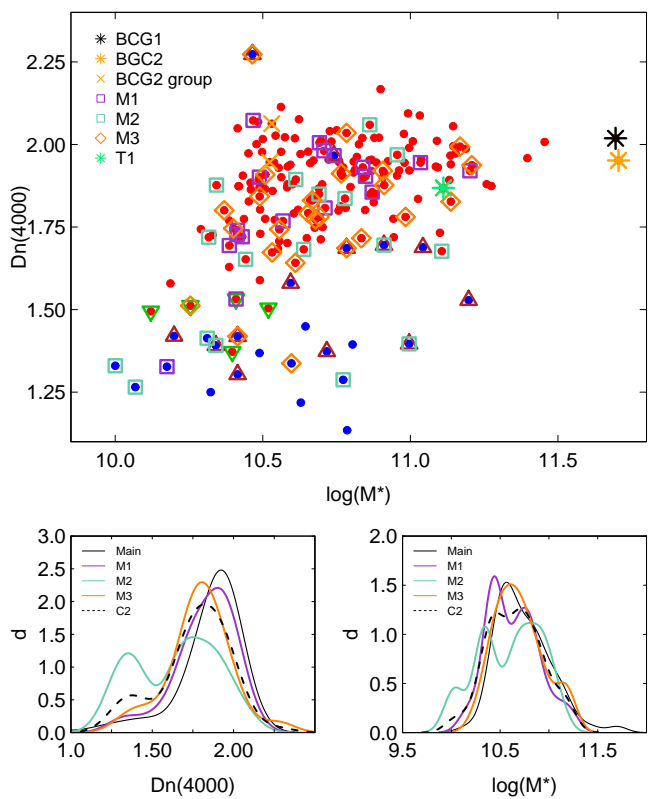}}
\caption{
Upper panel: $D_n(4000)$ index vs. stellar mass for the cluster A2142.
Red circles indicate low star formation rate galaxies with 
$\log \mathrm{SFR} < -0.5$, and blue circles shows high star formation rate
galaxies with $\log \mathrm{SFR} \geq -0.5$. Other notations are as in 
Fig.~\ref{fig:a2142radec}.
Lower panels: Probability density distributions of $D_n(4000)$ (left),
and stellar mass (right) of galaxies in the main cluster (black solid line), 
in the outskirts component
C2 (subclusters M1 - M3 taken together, black dashed line), and 
in subclusters M1--M3 (M1 - violet line, M2 - aquamarine line, and
M3 - orange line). 
}
\label{fig:a2142d4m}
\end{figure}

{\it Orientations and alignments}. 
We analysed the position angles in the sky plane of the whole cluster A2142 
and both BCGs,
and compared these  with the orientations of the X-ray and radio
haloes of the cluster \citep[see ][]{2013A&A...556A..44R,
2013ApJ...779..189F, 2016A&A...595A..42T, 2017A&A...603A.125V}.
We also found the position angle of the supercluster axis
determined by the distribution of galaxy groups with at least
10 member galaxies in the supercluster, 
and we found the position angles of subclusters $M1 - M3$. 
To determine the position angle of the supercluster, 
the cluster (including subclusters), 
and individual subclusters, we
approximated their shape with the ellipse and found the position angle of its
major axis.
We show the direction of the supercluster axis in 
Figs.~\ref{fig:radec} (black dashed line) and \ref{fig:axes} (black dotted line),
where we plot the sky distribution of galaxies 
in the cluster A2142 and in subclusters.
We also plot in Fig.~\ref{fig:axes} the directions of radio tails of galaxies T1, which coincides with the axis of subcluster M1, and T2 (pointing towards
M2) from \citet{2017A&A...603A.125V},
and approximate  the contours of the X-ray and radio haloes with ellipse
\citep{2017A&A...603A.125V}.

The position angles of the cluster
and supercluster axis approximately coincide, having values 
$50 \pm 3^{\circ}$ (A2142 cluster) and $63 \pm 1^{\circ}$ (supercluster axis),
measured counterclockwise
from west (see also Figs.~\ref{fig:radec} and \ref{fig:a2142radec}). 
Component C3 also lies at the supercluster axis.
Also the position angles of the visible major axis of both BGCs are close to these 
angles ($\approx 42^{\circ}$).
X-ray and radio haloes are aligned along the same axis 
\citep[Fig.~\ref{fig:axes} and ][]{2017A&A...603A.125V}.

The directions of tails of radio sources T1 and T2 \citep{2017A&A...603A.125V} 
coincide with the directions at which subclusters 
M1 and M2 are located. In the same time, both galaxies T1 and T2 are far from
subclusters M1 and M2, 
as they are located in the centre of the cluster in sky coordinates, 
at the projected distance
approximately $1.8 - 2$~\Mpc\ from subclusters. 
The direction of their tails may be just a coincidence. However,
\citet{2016ApJ...821...29L} have mentioned that the presence of the
tailed radio galaxy in A2142 may be explained by significant bulk motions
in the intercluster medium.
Thus the agreement of the directions of galaxy radio tails
with substructures suggest the possibility that the 
infall of subclusters
affect the properties of these galaxies.  

The subcluster M3 is positioned on the supercluster axis and its 
position angle is coincident with the supercluster axis
(Fig.~\ref{fig:axes}).
This agrees with simulations showing that infalling galaxy groups may be elongated along the direction 
of infall because they are stretched by the cluster tidal field \citep{2013MNRAS.435.2713V},
and supports the suggestion
that we evidence an infall of galaxies along this axis into the cluster.

\begin{figure}[ht]
\centering
\resizebox{0.45\textwidth}{!}{\includegraphics[angle=0]{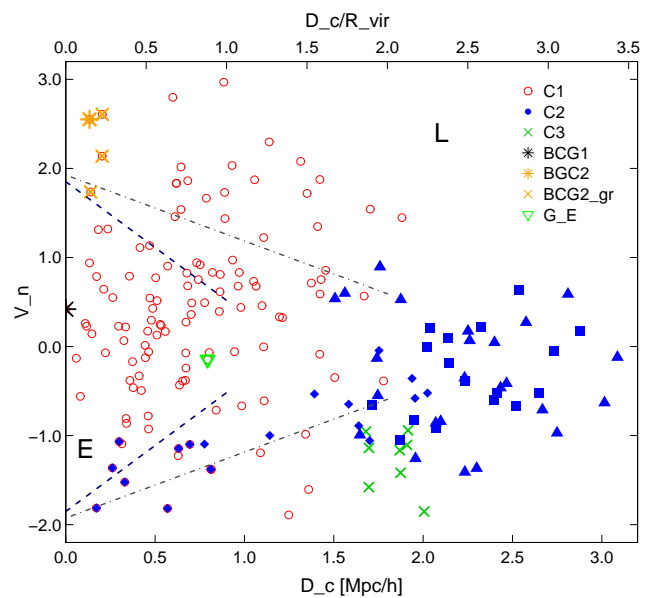}}
\caption{
Normalised velocity of galaxies with respect to the cluster mean velocity vs. 
projected clustercentric distance 
for the cluster A2142, its outskirts subclusters, and infalling groups (see text).
Lines separate approximately the early and late infall galaxies with increasing 
time of infall;  two different boundaries separate
early and late infall time galaxies.
The earliest infall galaxies with infall time 
$t_{\mathrm{inf}} > 1$~Gyr lie  between 
the black dashed lines, calculated using $R_{\mathrm{vir}}$,
and 
late or ongoing infall galaxies with infall time 
$t_{\mathrm{inf}} < 1$~Gyr lie  above and below
the lines, and on the right side of the lines 
\citep[see ][for details]{2013MNRAS.431.2307O, 2017MNRAS.467.4410A}.
The grey  dash-dotted lines are calculated using $R_{\mathrm{max}}$
(see text).
Notations of components C1--C3 and M1--M3
are the same as in Fig.~\ref{fig:a2142mclust}.
Notations of the BCGs and possible infalling groups are given
in legend;
E indicates the early infall region and L denotes the late infall region.
}
\label{fig:a2142cdvmclust}
\end{figure}

\begin{figure}[ht]
\centering
\resizebox{0.45\textwidth}{!}{\includegraphics[angle=0]{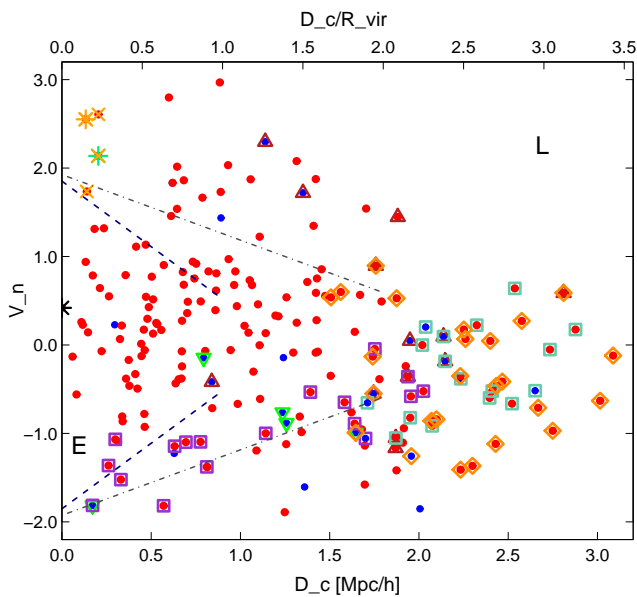}}
\caption{
Normalised velocity of galaxies with respect of the cluster mean velocity vs. 
projected clustercentric distance 
for the cluster A2142 and its outskirts.
Lines separate approximately the early and late infall galaxies with increasing 
time of infall; two different boundaries separate
early and late infall time galaxies.
Earliest infall galaxies with infall time 
$t_{\mathrm{inf}} > 1$~Gyr lie  between 
black dashed lines, calculated using $R_{\mathrm{vir}}$,
and 
late or ongoing infall galaxies with infall time 
$t_{\mathrm{inf}} < 1$~Gyr lie  above and below
the lines, and on the right side of the lines 
\citep[see ][for details]{2013MNRAS.431.2307O, 2017MNRAS.467.4410A}.
Grey  dash-dotted lines are calculated using $R_{\mathrm{max}}$.
Notations are the same as in Fig.~\ref{fig:a2142radec};
E indicates the early infall region and L indicates the late infall region.
}
\label{fig:a2142cdv}
\end{figure}

\subsection{Galaxy populations in the cluster A2142}
\label{sect:gal} 

Next we analyse galaxy populations in the cluster A2142 and its outskirts 
structures with the emphasis on their star formation histories.
To characterise galaxy populations at a glance,
we plot $D_n(4000)$ index versus stellar mass of galaxies
for galaxies from various 
structures determined in the cluster A2142 and its outskirts 
in upper panels of Fig.~\ref{fig:a2142d4m}. 
In lower panels of these figures, we plot the distributions 
of each parameter in the main cluster (component C1) and in subclusters
taken together (component C2), and separately in three subclusters M1 - M3.
We do not show error limits in the probability density distributions
in lower panels of Fig.~\ref{fig:a2142d4m} since Poisson errors are sensitive
to binning the data and may be misleading, especially for bimodal
and asymmetrical distributions. 
We used the Kolmogorov-Smirnov test to find the statistical significance 
of the differences in $D_n(4000)$ indexes for galaxies in the main cluster
and in outskirts (C2), using full data (the integral distributions). 
For details of this approach, we refer to \citet{2008ApJ...685...83E}.
We consider that the differences between distributions are highly significant 
if the $p$ - value (the estimated probability of rejecting the hypothesis
that distributions are statistically similar) $p \leq 0.01$.
  
Figure~\ref{fig:a2142d4m} shows that most galaxies in the main
cluster have old stellar
populations with no active star formation. In the main cluster
91\% of galaxies have $D_n(4000) \geq 1.55$.
The median value of the $D_n(4000)$ index for the main cluster is 
$D_n(4000)_{\mathrm{med}} = 1.9$. 
In the outskirts of component C2 the percentage of
galaxies with $D_n(4000) \geq 1.55$ is lower, 82\%,
and the median value of the $D_n(4000)$ index
$D_n(4000)_{\mathrm{med}} = 1.8$ (Fig.~\ref{fig:a2142d4m}).
The KS test showed that the differences between
$D_n(4000)$ index values are highly significant ($p < 0.01$).

The lower panels of Fig.~\ref{fig:a2142d4m} suggest that 
the distribution of $D_n(4000)$ indexes for 
galaxies in the subcluster M1 is similar to that for the 
main cluster galaxies.
In subcluster M3, $D_n(4000)_{\mathrm{med}} = 1.8$. 
The lower panels of the figure show that galaxies with  the same stellar
mass in the main cluster and in the subcluster M3 
have different distributions of 
$D_n(4000)$ indexes and therefore different star formation histories.
For very old stars this could also be different metallicities,
which have an effect on $D_n(4000)$ \citep{2003MNRAS.341...33K}. 
The subcluster M2 (Fig.~\ref{fig:a2142d4m}, lower left panel) 
embeds the largest percentage
of star-forming galaxies among subclusters. Here 35~\%
of  galaxies have young stellar populations and $D_n(4000)_{\mathrm{med}} = 1.7$.
This subcluster has the lowest stellar masses of galaxies
(Fig.~\ref{fig:a2142d4m}, lower right panel). According to 
\citet{2017A&A...605A...4H}, lower stellar mass galaxies also have
lower $D_n(4000)$ values. The different stellar mass distribution probably can 
explain the different $D_n(4000)$ distribution.
However, the number of galaxies 
in subclusters is rather small. This means that the differences mentioned here
have low statistical significance and can only be taken as suggestions. 

Galaxies from the possible BCG2 group (see Fig.~\ref{fig:a2142radec}) have as old stellar
populations as the main cluster galaxies, where the median value of
the $D_n(4000)$ index  $D_n(4000) = 1.9$. 
It is interesting to note that the galaxy identified with the 
radio source T1 from \citet{2017A&A...603A.125V} has the lowest
 $D_n(4000)$ index value among the galaxies from the possible BCG2 group.
The stellar age estimates of BGCs and T1 galaxy differ even more; 
they are $12.0$ and  $6.5$~Gyrs, respectively.
This gives additional   support to the suggestion that the
infall of galaxies from M1 subcluster 
affect the properties of this galaxy.  

We defined recently quenched galaxies as those
with $D_n(4000) \leq 1.55$ and star formation rate $\log \mathrm{SFR} < -0.5$ (five galaxies).
Red, high SFR galaxies are defined as galaxies with 
$g - r \geq 0.7$, and $\log \mathrm{SFR} \geq -0.5$ (13 galaxies).

Galaxies with low star formation rates and young stellar populations
(recently quenched galaxies with $D_n(4000) \leq 1.55$ 
and star formation rate $\log \mathrm{SFR} < -0.5$)
reside mostly in the transition region
 of Fig.~\ref{fig:a2142d4m}
and have intermediate $D_n(4000)$ index values. Most 
galaxies, which are red in colour and have high star formation rates
with $g - r \geq 0.7$, and $\log \mathrm{SFR} \geq -0.5$, also reside here.
Three recently quenched galaxies are located in the subcluster M3,
two of which belong to a close pair of galaxies. 
One recently quenched galaxy is located in the subcluster M1,
and one is located at the edge of the infalling group $G_E$ (Fig.~\ref{fig:a2142radec}).

\begin{figure*}[ht]
\centering
\resizebox{0.30\textwidth}{!}{\includegraphics[angle=0]{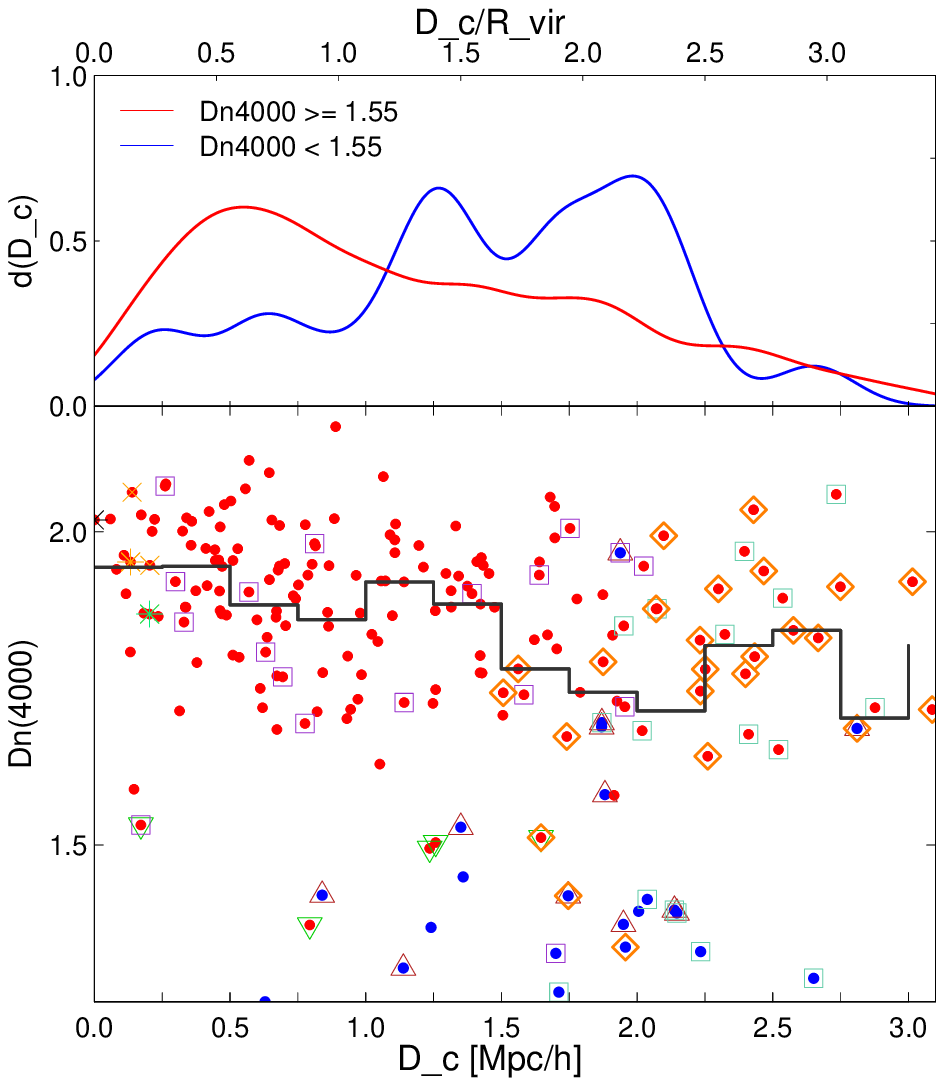}}
\resizebox{0.30\textwidth}{!}{\includegraphics[angle=0]{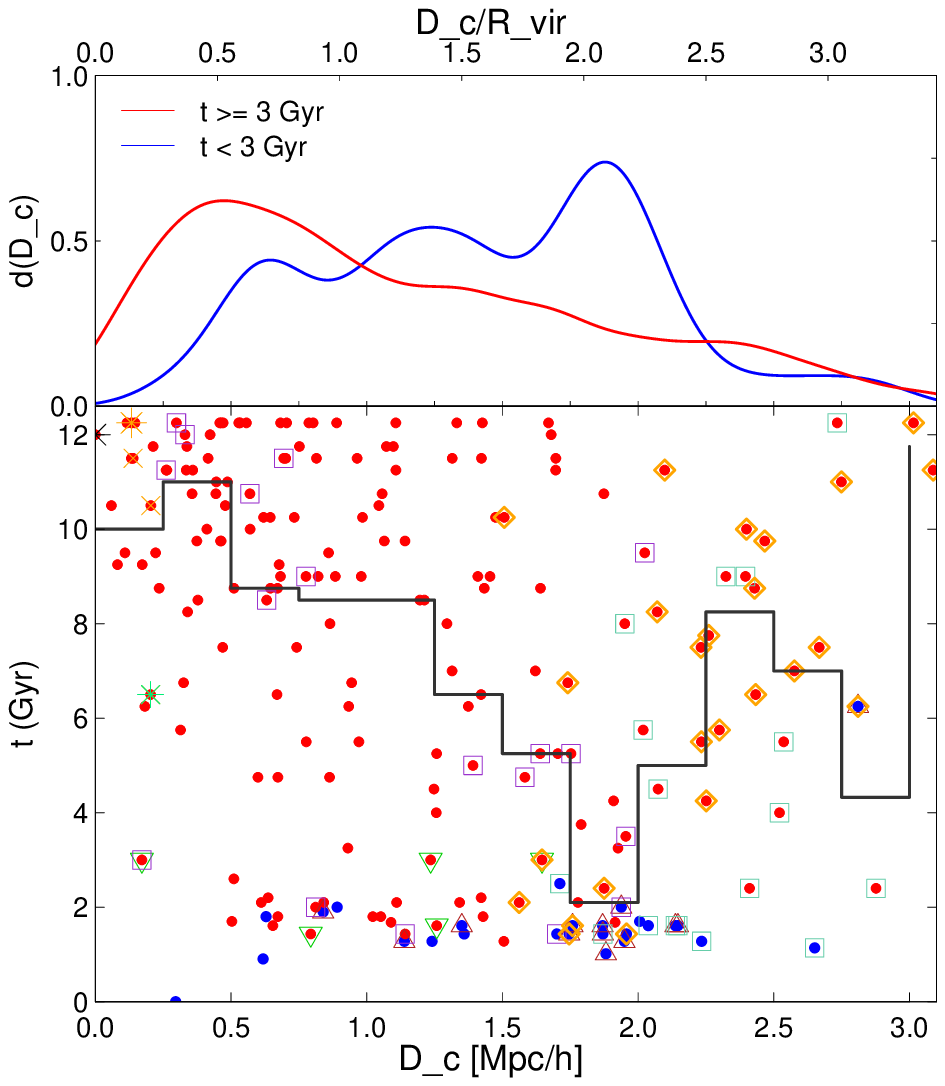}}
\resizebox{0.30\textwidth}{!}{\includegraphics[angle=0]{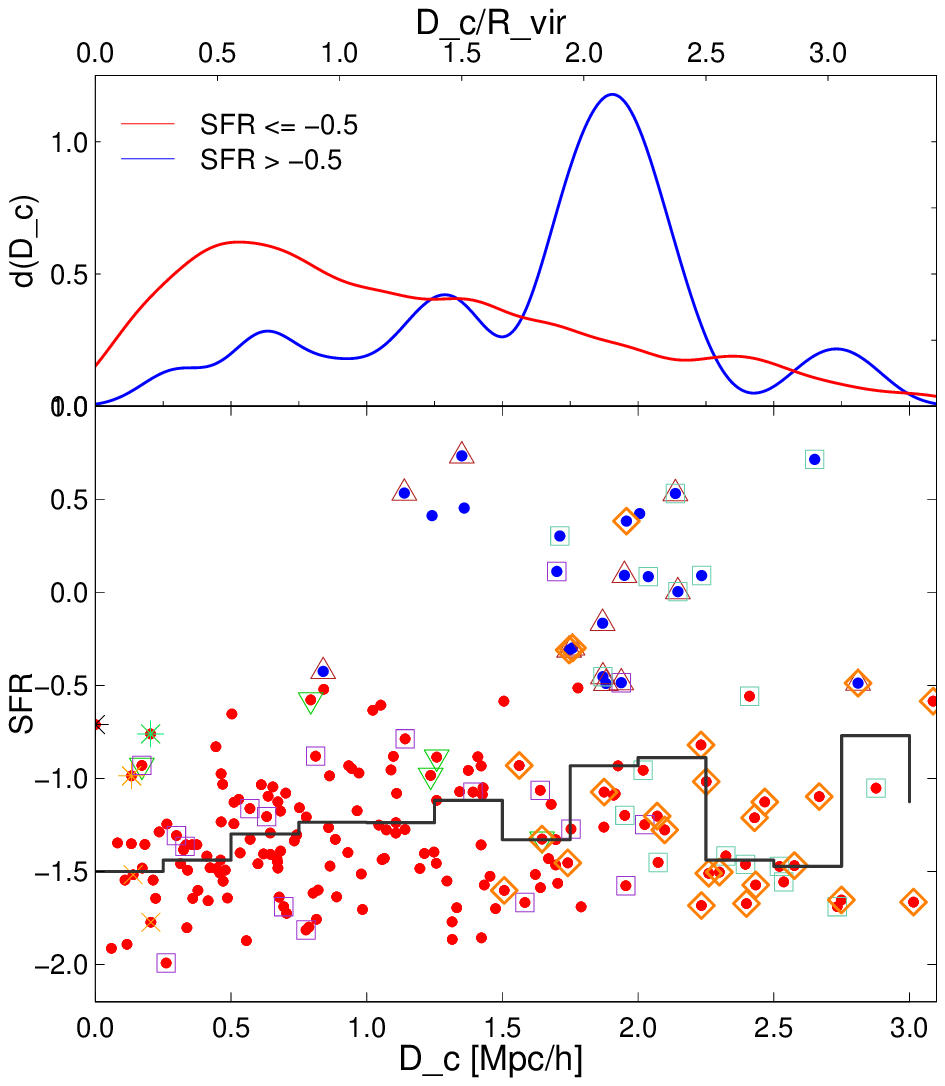}}
\caption{
$D_n(4000)$ index  (left panel),  ages of stellar populations (middle panel) 
and star formation rates (right panel)
for galaxies in A2142 vs. clustercentric distance.
The upper panels show the distribution of clustercentric distances 
for galaxies with young and old stellar populations, and passive and
actively star-forming galaxies (see legends in the panels).
In the lower panels, the black line shows median values of parameters.
Notations are as in Fig.~\ref{fig:a2142radec}, upper panel.
}
\label{fig:a2142cdpara}
\end{figure*}

\begin{figure*}[ht]
\centering
\resizebox{0.40\textwidth}{!}{\includegraphics[angle=0]{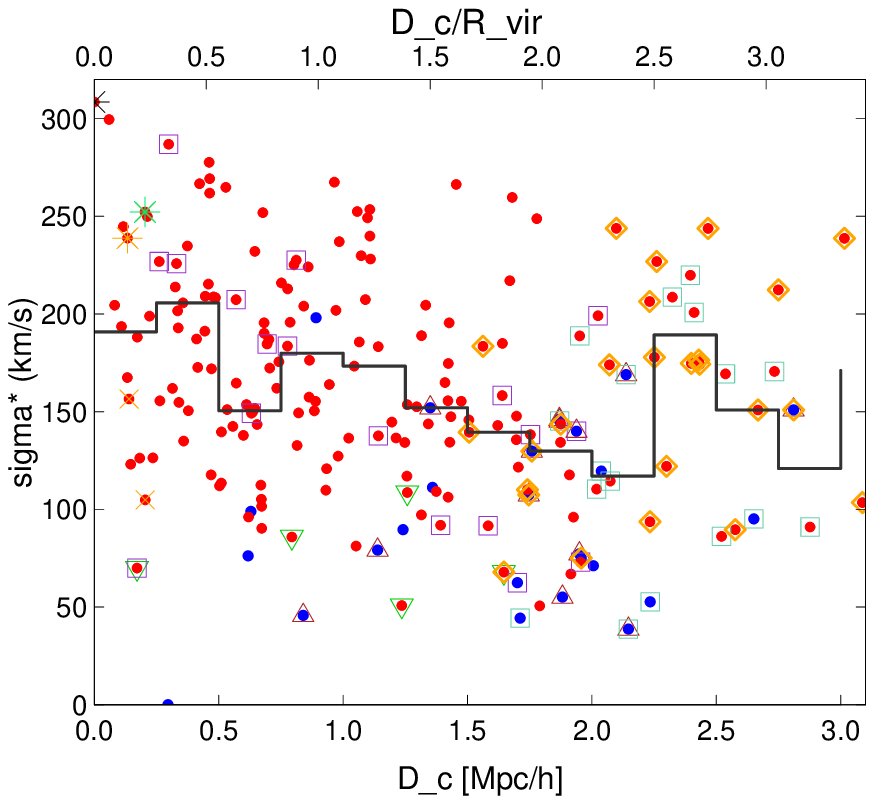}}
\resizebox{0.40\textwidth}{!}{\includegraphics[angle=0]{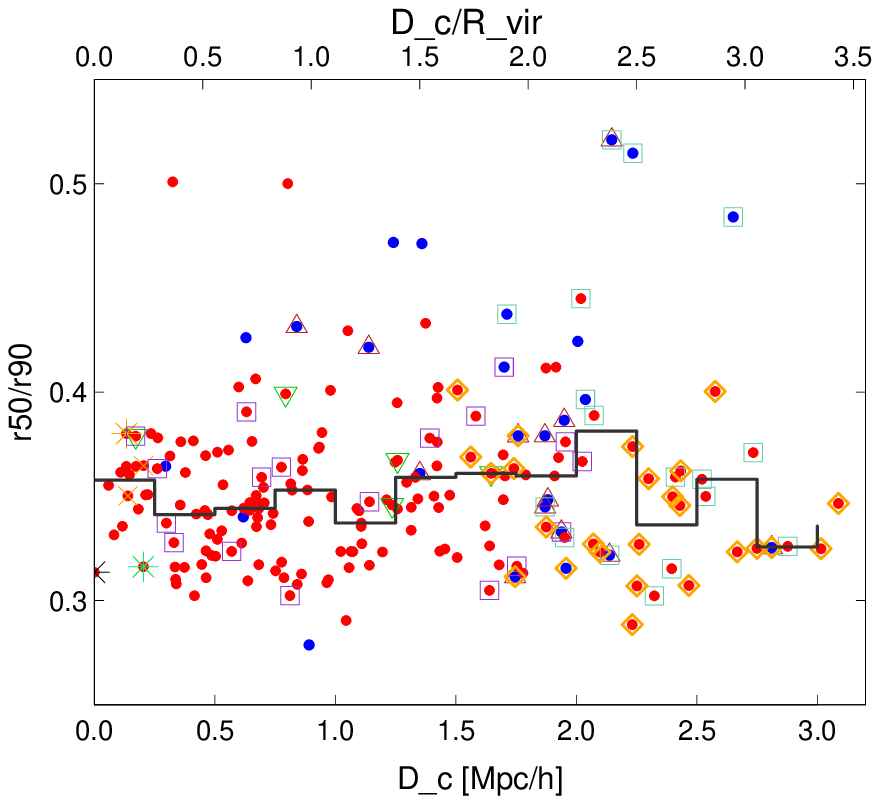}}
\caption{
Stellar velocity dispersions (left panel)
and concentration index (the ratio of the Petrosian radii 
$R_{50}/R_{90}$, right panel) for galaxies in A2142 vs. clustercentric distance.
The black line shows the median values of parameters.
Notations are as in Fig.~\ref{fig:a2142radec}, upper panel.
}
\label{fig:a2142cdsigma}
\end{figure*}

\subsection{Phase space analysis of the cluster A2142 and its outskirts}
\label{sect:phase} 

We employ projected phase space diagram of A2142 to separate regions of different
dynamical states and infall times and to analyse galaxy populations in these regions.
Corresponding PPS diagrams for the cluster A2142 are shown in
Figs.~\ref{fig:a2142cdvmclust} and \ref{fig:a2142cdv}.
The PPS diagrams show 
line-of-sight velocities of galaxies 
with respect  to the cluster mean velocity
versus projected clustercentric distance.
The differences between galaxy velocities
and the cluster mean velocity are normalised by the cluster velocity dispersion, 
$v_\mathrm{n} = (v - v_\mathrm{mean})/\sigma_\mathrm{cl}$.
Several recent studies have shown that in the phase space diagram galaxies with 
different accretion histories populate different areas of the diagram
\citep{2013MNRAS.431.2307O, 2014ApJ...796...65M, 2015ApJ...806..101H, 2015MNRAS.448.1715J,
2017MNRAS.467.4410A, 2017ApJ...838..148P, 2017ApJ...838...81Y, 2017ApJ...843..128R}. 
Galaxies with early infall times are located at small clustercentric distances
in the central part of the
figure (virialised region), while recently infallen or still infalling 
galaxies lie at large
clustercentric distances and/or have high velocities with respect of the
cluster mean velocity (nonvirialised region). 
Early and late infall time regions can be approximately 
separated with a line corresponding to the infall time of the cluster member galaxies
\citep{2013MNRAS.431.2307O} as follows:
\begin{equation}\label{eq:infalltime}
| \frac{v-v_\text{mean}}{\sigma_\text{cl}} |= 
-\frac{4}{3}\frac{D_\text{c}}{R_\text{vir}}+2 , 
\end{equation}
where $v$ are the velocities of the galaxies, 
$D_\text{c}$ is the projected clustercentric distance, and 
$R_{\mathrm{vir}}$ is the cluster virial radius. 
\citet{2013MNRAS.431.2307O} showed that in simulations a large percentage of 
galaxies  on the left of the line at small clustercentric 
radii form an early infall (virialised) population 
with $\tau_{\mathrm{inf}} > 1$~Gyr.
Galaxies 
on the right of this line at large clustercentric radii 
fell into the cluster during the last 1~Gyr. 
These galaxies form late or ongoing infall populations
with an infall time $\tau_{\mathrm{inf}} < 1$~Gyr. 
Simulations show that  galaxies with late infall time 
may populate areas of PPS diagram at all projected radii and 
velocities, which complicates the interpretation
of the PPS diagram \citep{2015ApJ...806..101H}.  

In Figs.~\ref{fig:a2142cdvmclust} and \ref{fig:a2142cdv} we use two lines to separate
different regions of A2142 in the PPS diagram.
Both are calculated with  Eq.~\ref{eq:infalltime}.
The dashed line corresponds to the virial radius as defined in
Eq.~\ref{eq:rvir}; it separates the inner, virialised  region of the cluster. 
The dot-dashed 
line corresponds to maximal radius of the components C1, 
$R_{\mathrm{max}}$. This line indicates the outer region 
of the main cluster where subclusters enter the cluster. 
We plot both set of lines to see how sensitive the PPS analysis is to the definition of the virialised region.

Figure~\ref{fig:a2142cdvmclust} presents the PPS diagram 
for substructures C1--C3 and subclusters of the component C2.
We also show here galaxies from the BCG2 group and possible $G_E$ group 
member. This figure shows that almost all  galaxies at clustercentric distances
approximately 
$D_c < 0.6$~\Mpc\  belong to the early infall (virialised) region
and  are members of the main component C1. 
Galaxies from the BCG2
group in the cluster centre in sky coordinates 
have very high velocities with respect to the cluster mean velocity
\citep[noted also in ][]{2000ApJ...541..542M}. 
These galaxies may have already passed
cluster centre \citep{2013MNRAS.431.2307O}. 
The galaxy at the edge of the group $G_E$ lies 
at the clustercentric distance approximately $0.7$~\Mpc, 
at the edge of the virialised region. 

The members of subclusters
M1--M3 and component C3 lie in the late infall region of the PPS diagram.
These members may still be infalling.
Also, some subcluster M1 galaxies, although near the cluster centre according to the
projected distance, 
lie in the region between two lines and may still be infalling.
\citet{2011ApJ...741..122O} detected several substructures 
in the direction of the subcluster M1  (their S5--S7). 
Infall of subcluster M1 galaxies in this region
may be the cause of the radio ridge described in  \citet{2017A&A...603A.125V}.

According to the position in the PPS diagram, the component C3
may represent another poor galaxy group infalling to the main cluster 
along the supercluster axis.

Figure~\ref{fig:a2142cdvmclust} also shows that some galaxies from the main cluster
component, C1, outside of the early infall region at clustercentric distances,
$D_c > 0.7$~\Mpc\ , have high positive velocities with respect to the cluster centre.
Simulations suggest  that such galaxies
may belong to a population yet to be accreted 
\citep[see e.g. Fig.~13 in][]{2015ApJ...806..101H}.

In Fig.~\ref{fig:a2142cdv} most galaxies of the main cluster
within the early infall region, including galaxies from the BCG2 group,  
 have old stellar populations ($D_n(4000) \geq 1.55$). 
One recently quenched galaxy and one red star-forming galaxy lie at the edge of the infalling
group $G_E$. 
 Their properties could be affected by the infalling group.

In Fig.~\ref{fig:a2142cdv}, star-forming (including red star-forming) 
and recently quenched  galaxies reside mostly in the region
at approximately $1.5 - 2.0$~\Mpc\ from the cluster centre at 
the outskirts  of C1 connected with subclusters M2 and M3, which are
probably infalling.  
 Galaxies from the poorest cluster component 
determined by {\it mclust} (one red and one blue  star-forming galaxy
among these members) are also located here. 
At clustercentric distances larger than $2$~\Mpc,
subclusters M1 and M3 are populated mostly by
passive, non-star-forming galaxies; star-forming galaxies belong mostly to
the subcluster M2.
We analyse how the properties of galaxies in A2142 and its outskirts 
change with the clustercentric distance in more detail in next subsection.

\subsection{Galaxy populations in A2142 versus projected clustercentric distance.}
\label{sect:dc} 

Next we study how the galaxy populations in the cluster and subclusters change
with the clustercentric distance with a focus on  star formation 
indicators. We calculate the distributions of
clustercentric distances for galaxies with various star formation
properties.  
We use parameter limits as given before. 
Quiescent galaxies with old stellar populations 
have  $D_n(4000) \geq 1.55$, and star-forming galaxies with young stellar 
populations have $D_n(4000) < 1.55$. 
Stellar age limit $t = 3$~Gyrs separates galaxies with old and young
stellar populations, galaxies with old stellar populations have $t \geq 3$~Gyrs,
and those with young stellar populations have $t < 3$~Gyrs.
Star formation rate $\log \mathrm{SFR} \leq -0.5$ corresponds to quiescent
galaxies, and $\log \mathrm{SFR} \geq -0.5$ to star-forming galaxies. 
In upper panels of Fig.~\ref{fig:a2142cdpara}, 
we show the distributions of clustercentric distances for galaxies
from these populations. Distributions are normalised, so that
each integrates to 1. In lower panels of this figure, we show the 
$D_n(4000)$ index, age, and star formation rates for all galaxies from
our sample. Additionally, in Fig.~\ref{fig:a2142cdsigma}
we show the stellar velocity dispersions 
and concentration index (the ratio of the Petrosian radii 
$R_{50}/R_{90}$) for galaxies in A2142 versus clustercentric distance. 
We also plot median values of the parameters for the full cluster.
In the upper panels, distributions are calculated for complete sample, 
excluding faint galaxies below the completeness limit, 
$M_r = -19.6$~$+5\log_{10}h$. 
The values of parameters are known for all galaxies (except one galaxy). 

The upper panels of Fig.~\ref{fig:a2142cdpara} show that the distribution
of passive, old galaxies without ongoing star formation has maximum at
clustercentric distances approximately $0.5$~\Mpc\  in the central,
most virialised  part of 
the main cluster. 
At higher clustercentric distances the percentage of 
passive galaxies decreases and star-forming galaxies become more frequent.
All panels of Fig.~\ref{fig:a2142cdpara} show that the maximum in the 
distance distribution of star-forming galaxies lie approximately at the distance
of $1.8$~\Mpc.  
The KS test shows that the differences between clustercentric distance
distributions of passive and star-forming galaxies have very high statistical
significance (with $p < 0.01$) for all three parameters. 

The lower panels of Fig.~\ref{fig:a2142cdpara} show that median value of the $D_n(4000)$ index 
for galaxies in the 
cluster centre at the clustercentric distances
$D_{c} < 0.5$~\Mpc\ is $D_n(4000) = 1.95$ and the median age
of stellar populations is $t \approx 10 - 11$~Gyrs.  
The galaxy T1 from a possible BCG2 group has the lowest
age among the BCG2 group galaxies and the highest star formation rate.

At the clustercentric distances
$D_{c} > 0.5$~\Mpc, median values of galaxy parameters start to change.
At the clustercentric distance interval  $1.75 < D_{c} < 2.25$~\Mpc,
at the 
boundaries of the main cluster where subclusters fall into the cluster
almost all galaxy parameters change rapidly. The 
age of stellar populations of galaxies have median value of $2$~Gyrs only, 
and the median value of the $D_n(4000)$ index is $D_n(4000) = 1.7$.
Stellar velocity dispersions of galaxies have the 
lowest median values and galaxies become less compact.
The recently quenched galaxies and  most star-forming galaxies
in subclusters M1, M2, and M3  lie in this distance interval. 
The KS test shows that the differences between the parameter values
in the centre of the cluster (at clustercentric distances
$D_{c} > 0.5$~\Mpc, 42 galaxies from the complete sample) and at the virial radius (in the 
clustercentric distance interval $1.75 < D_{c} < 2.25$~\Mpc, 33 galaxies)
are highly significant with $p < 0.01$.
 
Higher stellar velocity dispersions in the central part of the cluster,
in the region of early infall in the phase space diagram
(Fig.~\ref{fig:a2142cdv}) are in agreement with the results that
show that galaxies with old stellar populations have 
higher stellar velocity dispersion values than galaxies with young stellar
populations \citep{2009ApJ...694..867S, 2012ApJ...760...62B}.

As the clustercentric distance increases ($D_{c} > 2.25$~\Mpc),
stellar ages of galaxies increase together with the increase
of their $D_n(4000)$ index and stellar velocity dispersion.
The median age of stellar populations of galaxies in the distance interval
of $2.25 < D_{c} < 2.5$~\Mpc\ is $8$~Gyrs.
The concentration index of galaxies increases and then decreases again.
In this distance interval star-forming galaxies belong to subcluster M2.

Together these figures suggest that  in a narrow 
clustercentric distance interval, 
the star formation of galaxies in subclusters is enhanced at the boundaries of the cluster where subclusters enter the main cluster 
and is accompanied by structural changes
of galaxies and changes in other properties of galaxies. 
This suggest that the infall of galaxies into the cluster
enhances their star formation, also affecting their other properties.

\section{Discussion}
\label{sect:discussion} 

\subsection{Properties of galaxies at various clustercentric distances}
\label{sect:propvir} 
We found the change in galaxy populations in the cluster A2142
at the clustercentric distances of about
$1.6 - 1.8$~\Mpc,\ which is probably evidence of a merger-induced 
star formation activity in the region 
where subclusters fall into the main cluster.
This has also been found in other studies. For example, 
\citet{2008MNRAS.388.1152P} and \citet{2012MNRAS.427.1252M} detected enhanced
star formation in galaxies in filaments entering into galaxy clusters.
 \citet{2017arXiv170703208D} have
detected a high percentage of star-forming galaxies in the cluster A520
in infalling groups at $R_{200}$ and higher clustercentric distances.
\citet{2017ApJ...841...18B} have shown that colours of galaxies from the SDSS 
change at the halo boundaries.
\citet{2009ApJ...699.1595P} found that at the characteristic scale 
 approximately 
1 - 3 $R_{200}$ (for bright galaxies this radius is smaller than 
for faint galaxies) 
 the properties of galaxies start to depend on the clustercentric
radius at a fixed nearest neighbour environment. 
In the same time, some other studies detected smooth changes in galaxy
populations in clusters with the increase of clustercentric distance
\citep{2015ApJ...806..101H, 2016ApJ...816L..25P}. 
These studies propose that, for example, interactions of galaxies with intracluster 
medium causes a gradual shutdown of star formation.
Our results in Sect.~\ref{sect:dc} show both the slow changes in galaxy
properties as the clustercentric distance increases, and a 
rapid change in galaxy properties at the boundaries 
of the cluster. This may be related to accretion shocks
as gas falls into clusters along the filaments around clusters
\citep{2009ApJ...696.1640M}.

\subsection{Central region of A2142: Multiwavelength view}
\label{sect:central} 
Earlier studies of the central region of A2142 (see references in Introduction)
show that the central part of the cluster have a very complicated
structure, with many substructures, showing signatures of several mergers.
We described some of these merging structures
and their galaxy populations in this paper. 
Among these structures are infalling groups and subclusters. 
One recently quenched galaxy is found at the edge of infalling group
$G_E$ and is described in \citet{2014A&A...570A.119E} and \citet{2017A&A...605A..25E}. 
Stellar populations of galaxies from the possible BCG2 group are very old.
Infall of subcluster M1 galaxies 
may be the cause of the radio ridge described in  \citet{2017A&A...603A.125V}.
Also, 
\citet{2017A&A...603A.125V} have described two tailed radio galaxies 
in the central region of the cluster (T1 and T2). The directions of the
tails coincide with the direction at which subclusters 
M1 and M2 are located and supports the possibility that these subclusters
are infalling and affecting galaxy properties in the cluster.

These studies have also shown that  in this cluster the distribution of galaxies follow that of
intracluster medium. No signatures of the dark matter offset has been found as 
in some other rich merging galaxy clusters such as in the cluster A520 
\citep[examples of some such galaxy clusters
were given recently in ][]{2017arXiv170105877W}.
This can be explained by different timescales of mergers. For A520 the time
since merger has been estimated to be less than $0.5$~Gyr 
\citep{2007ApJ...668..806M, 2008A&A...491..379G, 2017arXiv170703208D},
while for A2142 we propose that the time since major merger 
is larger and may be of the order of $4$~Gyrs (see below).

\subsection{Clusters, superclusters, and the cosmic web}
\label{sect:multi}
In this study we showed that the cluster A2142, 
infalling subcluster M3, and one infalling group are aligned along the cluster and 
supercluster axis \citep[the cluster and supercluster alignments were also discussed
in][]{2015A&A...580A..69E}. 
The radio and X-ray haloes are also aligned  with the supercluster axis
\citep[][and Fig.~\ref{fig:axes}]{2017A&A...603A.125V}.

The correlated orientations of galaxies, 
galaxy clusters, and superclusters 
were noted already in the early studies of the cosmic web. 
\citet{1978MNRAS.185..357J} showed that in the Perseus--Pisces supercluster
the main galaxies of the clusters are directed along the chain that
connects the Perseus--Pisces supercluster with other nearby superclusters. 
Practically all clusters in the main chain of the Perseus–Pisces supercluster
are elongated along the main ridge of the chain. 
\citet{1978MNRAS.185..357J} concluded that a close physical link
exists between cluster main galaxies and their environment, which
hints at a common
origin and evolution of galaxies and galaxy clusters in the cosmic web.
Later many studies have confirmed the presence of alignments of galaxies
(especially the brightest cluster galaxies) 
in groups, clusters, superclusters, and filaments
\citep[][and references therein]{2000ApJ...543L..27W, 2002MNRAS.329L..47P, 
2005ApJ...618....1H,  2011MNRAS.414.2029P, 2015MNRAS.450.2727T,
2015A&A...576L...5T, 2016AAS...22723518M, 2017A&A...599A..31H, 2017A&A...601A.145F,
2017NatAs...1E.157W}.
The alignment signal may be a result of how galaxies and galaxy groups 
fall into clusters along preferred directions - 
the large-scale filamentary structures within which they are
embedded.

\citet{2012A&A...545A.104L} and \citet{2014A&A...562A..87E}
compared the percentage of red galaxies in
galaxy groups and clusters of similar richness 
in superclusters and in the field. These authors found that groups in superclusters
have higher 
percentage of red galaxies than groups in lower density
environment and that the  percentage of red galaxies is the 
highest in groups and clusters in the high-density cores of rich superclusters.

Our findings about the galaxy content of the main cluster A2142 and subclusters
M1 and M3 agree with the result by \citet{2012A&A...545A.104L} and \citet{2014A&A...562A..87E}.
Galaxies in the subcluster M1 have star formation properties closer to those 
in the main cluster than in other subclusters.
These galaxies lie in the outer region of the cluster
with later infall times than galaxies in the central, early infall region.
The percentage of star-forming galaxies in the subcluster M2 is higher than in M1 and M3; it is higher than typical for unrelaxed clusters, especially in supercluster
cores, $f_{SF} \approx 0.2$ \citep{2014ApJ...783..136C, 2017ApJ...835...56C}.
Also, \citet{2013MNRAS.432.1367L} found that galaxy populations
in groups residing in superclusters
are older than those in groups elsewhere. 
On the other hand, \citet{2010A&A...522A..92E} studied the substructure and 
galaxy content
of the richest galaxy clusters from the SGW. These authors used
{\it mclust} to determine different components in clusters. The analysis
of the galaxy content of individual components in clusters  showed that
in several clusters some components were populated mainly by elliptical
galaxies and other components by  spiral galaxies, 
which is in agreement with what we found for the cluster A2142 and subclusters M1 - M3.
The high percentage of star-forming galaxies in the subcluster M2 may be due to
anisotropic infall of galaxies such that, in this subcluster, galaxies
are affected by the infall up to higher distances from the cluster 
than in other subclusters.

\begin{figure}[ht]
\centering
\resizebox{0.44\textwidth}{!}{\includegraphics[angle=0]{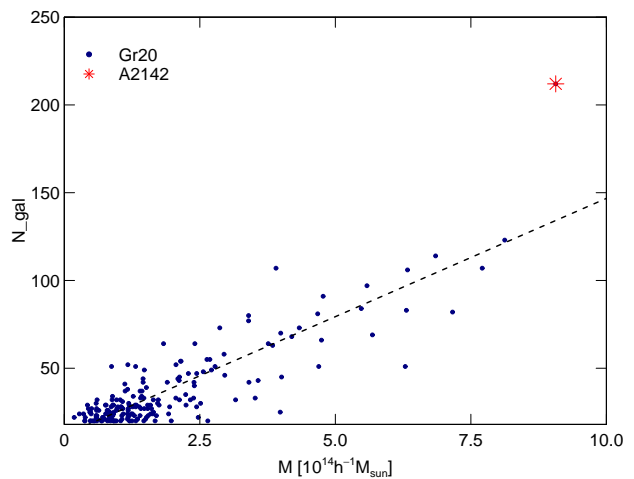}}
\caption{
Masses of galaxy groups with at least 20 member galaxies (dark blue symbols)
vs, the group richness 
in the  distance interval $225 - 280$~\Mpc\ from the group cataloque
by \citet{2014A&A...566A...1T}.
The red star denotes the cluster A2142.
}
\label{fig:massrichness}
\end{figure}

\begin{figure*}[ht]
\centering
\resizebox{0.95\textwidth}{!}{\includegraphics[angle=0]{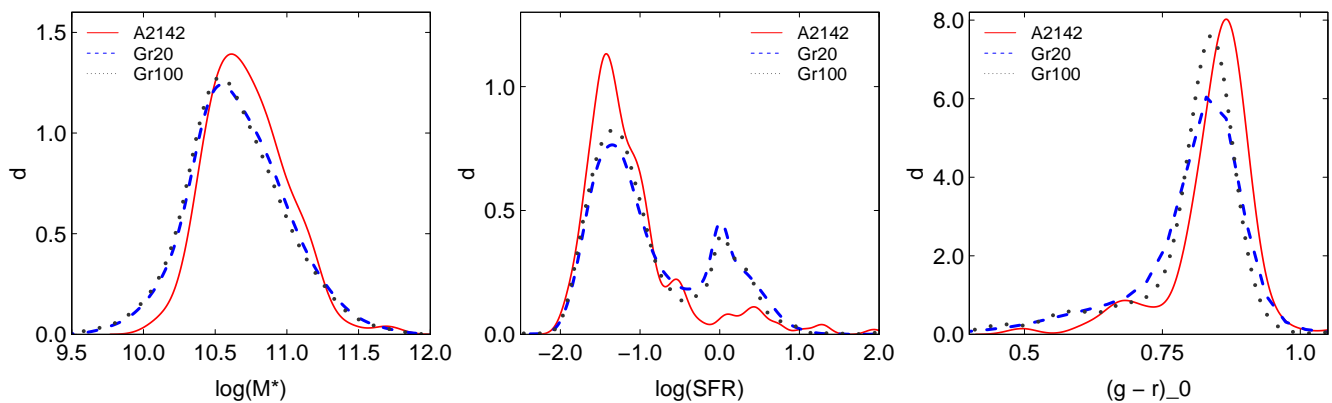}}
\caption{
Distributions of stellar masses (left panel), star formation rates (middle panel),
and $(g - r)_0$ colours (right panel) 
of galaxies in the cluster A2142 (red solid line), and in 
groups with at least 20 and 100 member galaxies 
(dashed blue and dotted grey lines, correspondingly)
in the  distance interval $225 - 280$~\Mpc\ from the group catalogue
by \citet{2014A&A...566A...1T}, for 
galaxies brighter than the completeness limit of A2142, 
$M_r = -19.6$~$+5\log_{10}h$.
}
\label{fig:grsmass}
\end{figure*}

\citet{2014A&A...562A..87E} analysed the structure and galaxy content
of galaxy superclusters from the SDSS and found that galaxy clusters in superclusters
with a more complicated inner structure (so-called superclusters of
spider morphology) have a higher percentage of blue star-forming galaxies 
than galaxy clusters in superclusters with simpler inner structures in which
galaxy groups and clusters are connected by a small number of galaxy
filaments (filament-type morphology). The A2142 supercluster is of 
filament morphology and the very high percentage of passive, red
galaxies in its main cluster agrees with the finding by \citet{2014A&A...562A..87E}.
In the same time we showed that the percentage of star-forming galaxies 
increases in the infall region of the cluster.
This may provide a hint as to why the percentage of star-forming galaxies 
is high in superclusters in which galaxy clusters are connected by a
large number of filaments (spider-type morphology). A large number of filaments
increases the possibility of mergers and increases star formation in galaxies.

\citet{2013MNRAS.428..906C} did not find a difference between 
stellar populations of galaxies in superclusters classified by their overall shape
as filaments and pancakes. In agreement with this result, \citet{2014A&A...562A..87E} 
found that superclusters with different overall shapes have similar  
galaxy populations. 

Individual rich superclusters have different inner structures and
galaxy and group contents
\citep{2014A&A...562A..87E}. For example, 
\citet{2013A&A...551A.143K} found that in the Ursa Major
supercluster groups with Gaussian galaxy velocity distributions reside
in higher density environments than groups with 
non-Gaussian galaxy velocity distributions and have higher densities of red galaxies.
The Ursa Major supercluster 
is classified as having multibranching filament type morphology 
in \citet{2011A&A...532A...5E}. Ursa Major has several rich clusters, while SCl~A2142
is dominated by one very rich cluster. 
However, the velocity distributions of galaxies in different components
in the cluster A2142 overlap, and using
velocity data only, it could be classified as  Gaussian \citep{2012A&A...540A.123E}.
This cluster resides in the high-density core  of the supercluster and has
high percentage of red galaxies, which agrees with the \citet{2013A&A...551A.143K}
findings for the Ursa Major supercluster.

\citet{2008MNRAS.390..289J} found 
an excess of  blue star-forming galaxies
aligned along the axis connecting the cluster A3158 to the cluster A3125/A3128
complex in the core region of the Horologium-Reticulum
supercluster. These authors concluded that this is evidence of merger-induced
star formation in these galaxies.

Superclusters obtained their 
group and galaxy populations
during the formation and evolution of the cosmic web. 
The observational results  show that
the variety of galaxy and group properties in individual 
superclusters is large.  This shows a need to continue the studies 
of galaxy and group populations in superclusters
to understand better their formation and evolution.

\subsection{Comparison with other galaxy groups and clusters}
\label{sect:SGW} 

Now we briefly compare A2142 with other rich groups and clusters
from the \citet{2014A&A...566A...1T} group catalogue.
To avoid strong  distance-dependent selection effects, 
we only used data 
 of  groups with at least 20 member galaxies 
from a narrow distance interval $225 - 280$~\Mpc.
Many rich superclusters including the SGW
are located at these distances \citep[see also][for details about 
galaxy superclusters from this distance interval]{2014A&A...562A..87E}.

Figure~\ref{fig:massrichness}  presents  the mass - richness relation
for galaxy groups. 
This figure shows that 
A2142 is much richer  than other groups and clusters in this distance interval.
For example, A2142 is twice as rich as the richest galaxy clusters in the SGW,  which is
a rich supercluster complex at approximately the same distance
as SCl~A2142 \citep[][and references therein]{2010A&A...522A..92E}.

\citet{2010A&A...522A..92E} studied substructure and
galaxy populations of the 10 richest galaxy clusters in the SGW superclusters.
These clusters show the presence of substructure, signs of mergers,  and perhaps
rotation, which suggest that the richest clusters in the SGW are not yet virialised. 
The mean values of the colour index  
of galaxies in these clusters, $(g - r)_{\mathrm{mean}} \approx 0.75$, are
lower than we found for the cluster A2142 and its substructures in this paper,  
$(g - r)_{\mathrm{med}} \approx 0.9$. \citet{2010A&A...522A..92E} also found
that the percentage of red galaxies in various SGW superclusters increases 
with supercluster richness and is higher in clusters that lie in the high-density
cores of the SGW superclusters. The  percentage of red galaxies is the highest 
in the clusters A1650 and A1750 in the cores of the richest SGW supercluster,
$f_{\mathrm{red}} \approx 0.9$, which is the same as in the main cluster in this study.
The masses of the richest clusters in the high-density cores of the SGW 
are approximately $M_{\mathrm{cl}} \approx  10^{15}h^{-1}M_\odot$, that is similar
to the mass of the cluster A2142 
\citep[Table~\ref{tab:gr10} and ][]{2016A&A...595A..70E}.

Next we compared galaxy populations in A2142 and in other
groups and clusters with at least 20 member galaxies
in the same distance interval as before, $225 - 280$~\Mpc. 
In the comparison  we only used galaxies  brighter than
$M_r = -19.6$~$+5\log_{10}h$, which is the completeness limit of A2142.
In Fig.~\ref{fig:grsmass} we plot distributions of stellar masses, 
star formation rates, and $(g - r)_0$ colours
of galaxies in groups with at least 20 member galaxies, and 
separately in rich clusters with at least 100 member galaxies. 

This figure shows that the galaxies in A2142 have 
higher stellar masses than galaxies of the same absolute magnitude limit
in other rich groups and clusters. Also, they have lower star formation rates
and redder colours, although, as we mentioned above, the percentage of red galaxies
in A2142 and in some rich clusters from the SGW is the same. 
The Kolmogorov-Smirnov test shows that the differences between galaxy parameters
in A2142 and other groups and clusters are statistically
significant at  high levels ($p$-value $p < 0.01$ for all parameters).

Therefore, A2142 not only has a much larger number of luminous galaxies than
other groups and clusters 
 in this distance interval, the galaxies 
in A2142 are different, on average, from galaxies in other groups and clusters.
They  are redder and have higher stellar masses and lower star formation rates.
Moreover, \citet{2016ApJ...827L...5M} showed that 
the abundance of member galaxies in haloes of virial masses comparable to that of A2142
from numerical simulations is significantly smaller than in A2142.

\subsection{Formation of A2142}
\label{sect:form}
The cluster A2142 is an unusually rich cluster in the collapsing core
of the supercluster, elongated along the supercluster axis
together with the subcluster M3 and component C3.
Numerical simulations show that  
high-density cores of superclusters are the locations  
where positive sections of density perturbations combine
and clusters form via merging and accretion of galaxy groups and clusters
\citep[][and references therein]{1996Natur.380..603B, 
2011A&A...531A.149S, 2012ARA&A..50..353K}.  
The cluster A2142 is the only rich cluster in the supercluster. We may assume that 
this cluster has been formed through merging and accretion of other groups and clusters in
the supercluster core region. 
This assumption is supported by the finding 
that the core region of the A2142 supercluster with radius of about $6 - 8$~\Mpc\
has reached turnaround and is collapsing
\citep{2015A&A...580A..69E, 2015A&A...581A.135G}.
In an expanding Universe 
turnaround radius is a radius at which matter stops expanding with the
Universe and starts collapsing. 
We may expect that galaxy systems surrounding the A2142 cluster up to
the turnaround radius in the supercluster core are infalling, which is 
in agreement with the results of this paper.
Along
the supercluster axis, the accretion of groups may be enhanced owing to high overall density and this may make the cluster unusually rich.

The mass of A2142 is of the order of $10^{15}h^{-1}M_\odot$ \citep{2014A&A...566A..68M,
2015A&A...580A..69E}.
\citet{2015JKAS...48..213K} analysed the evolution of halo mass along the 
major descendant trees in Horizon Run 4 simulations 
and showed that  massive halos with mass of the order of $10^{15}h^{-1}M_\odot$ 
have formed relatively recently. 
On average,  halos with a current mass of
$10^{15}h^{-1}M_\odot$  have
their half mass ($0.5\times~10^{15}h^{-1}M_\odot$) at $z = 0.5$ 
\citep[look-back time approximately $4$~Gyrs, 
see also][and references therein]{2017arXiv170904945H}. 

Ongoing and recent accretion of groups and subclusters increase
the mass of the cluster. 
Next we give mass estimates of infalling groups and subclusters
discussed in this study.

{\it BCG2 group.}
The second brightest galaxy in the cluster (BCG2) with high peculiar
velocity
and old stellar populations with $D_n(4000)_{\mathrm{med}} = 1.9$ and $g - r = 0.92$
may be the brightest galaxy of an infalling group or cluster.
We can use several
approaches to estimate the mass of a possible merging group. One possibility is to use the median mass estimate 
of poor groups of galaxies in the \citet{2014A&A...566A...1T}
catalogue, $M_{\mathrm{gr}} \approx  10^{13}h^{-1}M_\odot$. 
Assuming that this galaxy
is the brightest galaxy in a group or 
cluster we can estimate its mass using the
relation between the stellar mass $M_{\mbox{*}}$ of a galaxy and its halo mass 
$M_{\mbox{halo}}$ from \citet{2010ApJ...710..903M},
\begin{equation}
\frac{M_{*}}{M_{\mbox{halo}}}=2\left(\frac{M_{*}}{M_{\mbox{halo}}}\right)_0 
\left[\left(\frac{M_{\mbox{halo}}}{M_1}\right)^
{-\beta}+\left(\frac{M_{\mbox{halo}}}{M_1}\right)^\gamma\right]^{-1},
\label{eq:stmass}
\end{equation}
where $(M_{\mbox{*}}/M_{\mbox{halo}})_0=0.02817$ 
is the normalisation of the stellar to halo mass relation, 
the halo mass $M_{\mbox{halo}}$ is the virial mass of haloes, 
$M_1 = 7.925\times~10^{11}M_\odot$
is a characteristic mass,
and $\beta=1.068$ and $\gamma=0.611$ 
are the slopes of the low- and high-mass ends of the relation, respectively. 

Both BCGs have stellar masses of about $M_{\mathrm{*}} \approx  5\times~10^{11}h^{-1}M_\odot$,
and we obtain that they could have haloes with mass of the order of
$M_{\mathrm{halo}} \approx  10^{14}h^{-1}M_\odot$. 
This is lower mass limit only and the actual halo masses may be higher
\citep{2017arXiv170505373M}. 
Several mass estimates for the cluster A2142 give mass of the order of  
$M \approx  10^{15}h^{-1}M_\odot$ \citep{2014A&A...566A..68M, 2015A&A...580A..69E}.
Therefore the range
of mass ratios of the infalling group and main cluster from different estimates 
is wide. However, our estimates are of the same order as obtained in 
\citet{2013A&A...556A..44R} who mention that the present data do not yet constrain this mass
very strongly. We note that galaxies from the possible BCG2 group have
stellar populations characteristic of the central parts of rich clusters,
supporting the possibility that they belonged to the rich cluster before merging.
 
{\it $G_E$ group.}
In Fig.~\ref{fig:a2142radec} we identified galaxies located at the edge of the
infalling group $G_E$, which is described in \citet{2014A&A...570A.119E} and in
\citet{2017A&A...605A..25E}.
Infall of the group affects its gas and may induce recent 
star formation activity in galaxies that are 
now observed as recently quenched and have young stellar populations with stellar ages
of approximately $2$~Gyrs (Fig.~\ref{fig:a2142cdpara}). 
This coincides with the estimate of a timescale of a mixing of the infalling
group gas with the surrounding intracluster medium \citep{2017A&A...605A..25E}. 
In the projected phase space diagram, galaxies from this group lie at early infall region
with infall time larger than $1$~Gyrs (Fig.~\ref{fig:a2142cdv}), which agrees
with the timescale for infall.
Typical mass of poor groups of galaxies is of the order of  $M_{\mathrm{gr}} \approx  10^{13}h^{-1}M_\odot$;
this coincides with the mass estimation for this group by \citet{2014A&A...570A.119E}.

{\it C3 group.} The poorest component in A2142 
may represent  galaxy group infalling to the main cluster 
along the supercluster axis.
The presence of star-forming galaxies in this group
may be an indication of star formation induced by infall.
\citet{2015A&A...580A..69E} found that its mass is of the order of
$M \approx 1.5\times~10^{14}h^{-1}M_\odot$. 
If we estimate its mass using the stellar mass of its brightest galaxy
we get a mass estimate of the group of approximately  $M \approx 1\times~10^{14}h^{-1}M_\odot$,
which is comparable to the mass estimate for the BCG2 group.

{\it Subclusters M1 - M3.}
The mass of the cluster A2142 
outskirts component $C2$ (all three subclusters together)
is approximately $M_{\mathrm{C2}} \approx 2\times~10^{14}h^{-1}M_\odot$ 
\citep[Table~\ref{tab:gr10} and ][]{2015A&A...580A..69E}.
To estimate masses of individual subclusters, we cannot apply the relation between
stellar mass and halo mass, as above for the possible BCG groups, since this relation
is derived for the central galaxies in virialised haloes \citep{2010ApJ...710..903M}. 
Therefore we used a very simple mass estimate.
Stellar mass in galaxies form approximately 1\% of the supercluster mass in which
they are embedded \citep{2015A&A...580A..69E, 2016A&A...595A..70E}.
If we consider that the stellar mass of galaxies in subclusters M1 - M3
gives~ 1 \% of their total mass, then we obtain from the sum of stellar
masses of galaxies in subclusters  that the mass
of the subclusters M1 and M2 is  $M \approx 0.9\times~10^{14}h^{-1}M_\odot$
and the mass of the subcluster M3 is $M \approx 1.5\times~10^{14}h^{-1}M_\odot$.
For the full outskirts region these estimates give mass on the same order as
the estimate for $C2$ in \citet{2015A&A...580A..69E}.

Together these estimates give the total mass in infalling groups and subclusters
approximately $M \approx 6\times~10^{14}h^{-1}M_\odot$.  \citet{2015JKAS...48..213K} 
predicted from the analysis of
the mass growth of massive haloes in Horizon Run 4 simulations that
haloes with mass of the order of $10^{15}h^{-1}M_\odot$ at redshift $z = 0,$
similar to cluster A2142, have gathered half of their mass during last $4$~Gyrs. 
Therefore the total mass in infalling subclusters is enough 
for the mass growth of A2142 during this period.  
However, simulations also predict that up to 40\% of mass growth of
haloes comes from smooth accretion of dark matter that was never bound in smaller 
haloes \citep{2010ApJ...719..229G, 2017A&A...605A...4H}. 
This suggests that  we may have overestimated the mass of infalling subclusters.

At the same time, A2142 embeds a surprisingly large number of luminous galaxies for
its richness (Fig.~\ref{fig:massrichness}).
A large number of luminous galaxies may come from a formation of
the cluster by merging two or more rich groups and clusters. 
The magnitude gap of the brightest galaxies in clusters
is one indicator of the time since the last major merger
\citep[][and references therein]{2017MNRAS.472.3246M}.
In A2142, the BCG2 is associated with the infalling group;
the last major merger may be related to the third brightest galaxy, BCG3.
In A2142 the magnitude gap between
the BCG1 and BCG3 in $r$-band
is one magnitude, the BCG3 lies at the projected clustercentric distance of
 $D_c \approx 0.45$~\Mpc. This is in agreement with the suggestion
 that the time since the last major merger in A2142 may be at least $4$~Gyrs  
\citep{2013ApJ...777..154D, 2017MNRAS.472.3246M}.

This prediction supports the suggestion that the possible major merger of rich
clusters, which could be the A2142 progenitors, could have happened
not later than $4$~Gyrs ago. 
In the projected phase space diagram  galaxies with infall time of the order of 
$4$~Gyrs occupy an early infall region
with clustercentric distances less than  $0.5$~\Mpc\ \citep{2015ApJ...806..101H}.
In A2142 this region is populated by galaxies with very old stellar populations
with median age of about $10 - 11$~Gyrs.
This suggests that galaxy populations in the central region of the cluster
may belong to the progenitors
of A2142 cluster before accreting. 
In addition, \citet{2014ApJ...786...79M} used simulations to show that
the brightest galaxies with  large peculiar velocities are 
associated with major mergers between clusters. If
mergers of clusters took place at latest at redhifts $z \geq 0.3,$
then clusters had time to reach equilibrium at present. However, if
clusters are constantly disturbed by mergers they may not reach equilibrium,
as in the case of A2142 where we are witnessing ongoing or recent mergers
by groups and subclusters of galaxies causing gas sloshing in the cluster.

The presence of substructure in A2142 may influence mass estimations, but
\citet{2017arXiv170910108O} have showed that although the mass estimates of galaxy clusters
may be biased in the presence of substructures, 
the most massive clusters are largely unaffected.
Also, various mass estimates
of A2142 agree well \citep{2014A&A...566A..68M, 2015A&A...580A..69E, 2017A&A...605A..25E}.
Therefore, such mergers may explain the richness of A2142 
but the discrepancy of mass remains.

\section{Summary}
\label{sect:sum} 

The analysis of the structure and galaxy populations
in the cluster A2142 and its outskirts revealed several infalling
groups and subclusters and showed a rapid change in galaxy 
properties 
in the infall region of the cluster at
clustercentric distances $1.6 - 2.2$~\Mpc. 
The central, virialised region  of the cluster at
clustercentric distances $D_c < 0.5$~\Mpc\ is populated with
galaxies with old stellar population that have ages of $10 - 11 $~Gyrs. 
Even in this region the properties of galaxies are affected by recent mergers.
With the increase of the 
clustercentric distance the properties of galaxies
change and  in the infall region of the 
cluster at $D_c \approx 1.6 - 1.8$~\Mpc\ 
median stellar ages of galaxies are approximately  $2$~Gyrs only.
Most star-forming or recently quenched galaxies lie in this region.
At higher
clustercentric distances ($D_c > 2.25$~\Mpc) the median age of galaxies is  
approximately $8$~Gyrs, which is lower than in the cluster centre
(although the scatter in ages is large). At this distance interval
star-forming galaxies lie in subcluster M2.

Infalling groups and subclusters have different
galaxy contents. One of these (M1) may cause radio ridge of A2142.
Cluster BCGs, subcluster M3,  and the cluster itself
are aligned along the supercluster axis. 
Ongoing mergers by infalling groups and subclusters 
give rise to the X-ray and radio structures of the cluster and affect the properties of
galaxies in the cluster. 
We speculate that A2142 has been formed through merging and accretion 
of other groups and clusters at least $4$~Gyrs ago along
the supercluster axis where the accretion of groups may be enhanced due to
high overall density.
Subcluster M3, infalling along the supercluster axis, may represent 
another merging cluster along the supercluster axis. 
The estimated total mass in infalling groups and subclusters, 
$M \approx 6\times~10^{14}h^{-1}M_\odot$. To be consistent with the
predicted mass
growth of haloes with mass of the order of $10^{15}h^{-1}M_\odot$ 
from redshift $z = 0.5$ (half-mass epoch) to the present in  
simulations we assume that we may have overestimated the mass in subclusters.  
Stellar masses of galaxies in A2142 are, on average, higher, their star formation rates are lower, and colours 
are redder than those in other groups
and clusters.
The richness of the cluster A2142 is also higher than the richness of other 
rich clusters of the same mass.
The reasons for these differences are not yet clear.
Clarifying this 
may lead to better understanding about
the formation and evolution of rich galaxy clusters
and the properties of dark matter.

In a forthcoming  work, we analyse the whole SCl~A2142, its galaxy, group, and filament
populations in the collapsing core and in outer regions to study
the complex evolution of galaxies in the cluster and in the whole supercluster.
We also continue to study rich, merging galaxy clusters and their galaxy populations
in collapsing cores of superclusters to understand better their coevolution 
within the cosmic web.

\begin{acknowledgements}

 We thank the referee for suggestions and comments that helped improve the paper. 
We are pleased to thank the SDSS Team for the publicly available data
releases.  Funding for the Sloan Digital Sky Survey (SDSS) and SDSS-II has been
provided by the Alfred P. Sloan Foundation, the Participating Institutions,
the National Science Foundation, the U.S.  Department of Energy, the
National Aeronautics and Space Administration, the Japanese Monbukagakusho,
and the Max Planck Society, and the Higher Education Funding Council for
England.  The SDSS website is \texttt{http://www.sdss.org/}.
The SDSS is managed by the Astrophysical Research Consortium (ARC) for the
Participating Institutions.  The Participating Institutions are the American
Museum of Natural History, Astrophysical Institute Potsdam, University of
Basel, University of Cambridge, Case Western Reserve University, The
University of Chicago, Drexel University, Fermilab, the Institute for
Advanced Study, the Japan Participation Group, The Johns Hopkins University,
the Joint Institute for Nuclear Astrophysics, the Kavli Institute for
Particle Astrophysics and Cosmology, the Korean Scientist Group, the Chinese
Academy of Sciences (LAMOST), Los Alamos National Laboratory, the
Max-Planck-Institute for Astronomy (MPIA), the Max-Planck-Institute for
Astrophysics (MPA), New Mexico State University, Ohio State University,
University of Pittsburgh, University of Portsmouth, Princeton University,
the United States Naval Observatory, and the University of Washington.

The present study was supported by the ETAG projects 
IUT26-2 and IUT40-2, and by the European Structural Funds
grant for the Centre of Excellence "Dark Matter in (Astro)particle Physics and
Cosmology" TK133. HL is funded by PUT1627 grant from Estonian Research Council.
This work has also been supported by
ICRAnet through a professorship for Jaan Einasto.
We thank Prof. Juhan Kim  for comments and Jukka Nevalainen
for inspiring discussions.

\end{acknowledgements}

\bibliographystyle{aa}
\bibliography{a2142gal.bib}

\begin{thebibliography}{111}
\expandafter\ifx\csname natexlab\endcsname\relax\def\natexlab#1{#1}\fi

\bibitem[{{Agulli} {et~al.}(2017){Agulli}, {Aguerri}, {Diaferio}, {Dominguez
  Palmero}, \& {S{\'a}nchez-Janssen}}]{2017MNRAS.467.4410A}
{Agulli}, I., {Aguerri}, J.~A.~L., {Diaferio}, A., {Dominguez Palmero}, L., \&
  {S{\'a}nchez-Janssen}, R. 2017, \mnras, 467, 4410

\bibitem[{{Ahn} {et~al.}(2014){Ahn}, {Alexandroff}, {Allende Prieto}, {Anders},
  {Anderson}, {Anderton}, {Andrews}, {Aubourg}, {Bailey}, {Bastien}, \&
  et~al.}]{2014ApJS..211...17A}
{Ahn}, C.~P., {Alexandroff}, R., {Allende Prieto}, C., {et~al.} 2014, \apjs,
  211, 17

\bibitem[{{Aihara} {et~al.}(2011){Aihara}, {Allende Prieto}, {An}, {Anderson},
  {Aubourg}, {Balbinot}, {Beers}, {Berlind}, {Bickerton}, {Bizyaev}, {Blanton},
  {Bochanski}, {Bolton}, {Bovy}, {Brandt}, {Brinkmann}, {Brown}, {Brownstein},
  {Busca}, {Campbell}, {Carr}, {Chen}, {Chiappini}, {Comparat}, {Connolly},
  {Cortes}, {Croft}, {Cuesta}, {da Costa}, {Davenport}, {Dawson}, {Dhital},
  {Ealet}, {Ebelke}, {Edmondson}, {Eisenstein}, {Escoffier}, {Esposito},
  {Evans}, {Fan}, {Femen{\'{\i}}a Castell{\'a}}, {Font-Ribera}, {Frinchaboy},
  {Ge}, {Gillespie}, {Gilmore}, {Gonz{\'a}lez Hern{\'a}ndez}, {Gott}, {Gould},
  {Grebel}, {Gunn}, {Hamilton}, {Harding}, {Harris}, {Hawley}, {Hearty}, {Ho},
  {Hogg}, {Holtzman}, {Honscheid}, {Inada}, {Ivans}, {Jiang}, {Johnson},
  {Jordan}, {Jordan}, {Kazin}, {Kirkby}, {Klaene}, {Knapp}, {Kneib},
  {Kochanek}, {Koesterke}, {Kollmeier}, {Kron}, {Lampeitl}, {Lang}, {Le Goff},
  {Lee}, {Lin}, {Long}, {Loomis}, {Lucatello}, {Lundgren}, {Lupton}, {Ma},
  {MacDonald}, {Mahadevan}, {Maia}, {Makler}, {Malanushenko}, {Malanushenko},
  {Mandelbaum}, {Maraston}, {Margala}, {Masters}, {McBride}, {McGehee},
  {McGreer}, {M{\'e}nard}, {Miralda-Escud{\'e}}, {Morrison}, {Mullally},
  {Muna}, {Munn}, {Murayama}, {Myers}, {Naugle}, {Fausti Neto}, {Cuong Nguyen},
  {Nichol}, {O'Connell}, {Ogando}, {Olmstead}, {Oravetz}, {Padmanabhan},
  {Palanque-Delabrouille}, {Pan}, {Pandey}, {P{\^a}ris}, {Percival},
  {Petitjean}, {Pfaffenberger}, {Pforr}, {Phleps}, {Pichon}, {Pieri}, {Prada},
  {Price-Whelan}, {Raddick}, {Ramos}, {Reyl{\'e}}, {Rich}, {Richards}, {Rix},
  {Robin}, {Rocha-Pinto}, {Rockosi}, {Roe}, {Rollinde}, {Ross}, {Ross},
  {Rossetto}, {S{\'a}nchez}, {Sayres}, {Schlegel}, {Schlesinger}, {Schmidt},
  {Schneider}, {Sheldon}, {Shu}, {Simmerer}, {Simmons}, {Sivarani}, {Snedden},
  {Sobeck}, {Steinmetz}, {Strauss}, {Szalay}, {Tanaka}, {Thakar}, {Thomas},
  {Tinker}, {Tofflemire}, {Tojeiro}, {Tremonti}, {Vandenberg}, {Vargas
  Maga{\~n}a}, {Verde}, {Vogt}, {Wake}, {Wang}, {Weaver}, {Weinberg}, {White},
  {White}, {Yanny}, {Yasuda}, {Yeche}, \& {Zehavi}}]{2011ApJS..193...29A}
{Aihara}, H., {Allende Prieto}, C., {An}, D., {et~al.} 2011, \apjs, 193, 29

\bibitem[{{Balogh} {et~al.}(1999){Balogh}, {Morris}, {Yee}, {Carlberg}, \&
  {Ellingson}}]{1999ApJ...527...54B}
{Balogh}, M.~L., {Morris}, S.~L., {Yee}, H.~K.~C., {Carlberg}, R.~G., \&
  {Ellingson}, E. 1999, \apj, 527, 54

\bibitem[{{Baxter} {et~al.}(2017){Baxter}, {Chang}, {Jain}, {Adhikari},
  {Dalal}, {Kravtsov}, {More}, {Rozo}, {Rykoff}, \&
  {Sheth}}]{2017ApJ...841...18B}
{Baxter}, E., {Chang}, C., {Jain}, B., {et~al.} 2017, \apj, 841, 18

\bibitem[{{Bezanson} {et~al.}(2012){Bezanson}, {van Dokkum}, \&
  {Franx}}]{2012ApJ...760...62B}
{Bezanson}, R., {van Dokkum}, P., \& {Franx}, M. 2012, \apj, 760, 62

\bibitem[{{Blanton} {et~al.}(2003){Blanton}, {Hogg}, {Bahcall}, {Brinkmann},
  {Britton}, {Connolly}, {Csabai}, {Fukugita}, {Loveday}, {Meiksin}, {Munn},
  {Nichol}, {Okamura}, {Quinn}, {Schneider}, {Shimasaku}, {Strauss}, {Tegmark},
  {Vogeley}, \& {Weinberg}}]{2003ApJ...592..819B}
{Blanton}, M.~R., {Hogg}, D.~W., {Bahcall}, N.~A., {et~al.} 2003, \apj, 592,
  819

\bibitem[{{Blanton} \& {Roweis}(2007)}]{2007AJ....133..734B}
{Blanton}, M.~R. \& {Roweis}, S. 2007, \aj, 133, 734

\bibitem[{{Bond} {et~al.}(1996){Bond}, {Kofman}, \&
  {Pogosyan}}]{1996Natur.380..603B}
{Bond}, J.~R., {Kofman}, L., \& {Pogosyan}, D. 1996, \nat, 380, 603

\bibitem[{{Brinchmann} {et~al.}(2004){Brinchmann}, {Charlot}, {White},
  {Tremonti}, {Kauffmann}, {Heckman}, \& {Brinkmann}}]{2004MNRAS.351.1151B}
{Brinchmann}, J., {Charlot}, S., {White}, S.~D.~M., {et~al.} 2004, \mnras, 351,
  1151

\bibitem[{{Bruzual} \& {Charlot}(2003)}]{2003MNRAS.344.1000B}
{Bruzual}, G. \& {Charlot}, S. 2003, \mnras, 344, 1000

\bibitem[{{Cappellari} \& {Emsellem}(2004)}]{2004PASP..116..138C}
{Cappellari}, M. \& {Emsellem}, E. 2004, \pasp, 116, 138

\bibitem[{{Chon} {et~al.}(2015){Chon}, {B{\"o}hringer}, \&
  {Zaroubi}}]{2015A&A...575L..14C}
{Chon}, G., {B{\"o}hringer}, H., \& {Zaroubi}, S. 2015, \aap, 575, L14

\bibitem[{{Cohen} {et~al.}(2014){Cohen}, {Hickox}, {Wegner}, {Einasto}, \&
  {Vennik}}]{2014ApJ...783..136C}
{Cohen}, S.~A., {Hickox}, R.~C., {Wegner}, G.~A., {Einasto}, M., \& {Vennik},
  J. 2014, \apj, 783, 136

\bibitem[{{Cohen} {et~al.}(2017){Cohen}, {Hickox}, {Wegner}, {Einasto}, \&
  {Vennik}}]{2017ApJ...835...56C}
{Cohen}, S.~A., {Hickox}, R.~C., {Wegner}, G.~A., {Einasto}, M., \& {Vennik},
  J. 2017, \apj, 835, 56

\bibitem[{{Costa-Duarte} {et~al.}(2013){Costa-Duarte}, {Sodr{\'e}}, \&
  {Durret}}]{2013MNRAS.428..906C}
{Costa-Duarte}, M.~V., {Sodr{\'e}}, L., \& {Durret}, F. 2013, \mnras, 428, 906

\bibitem[{{Deason} {et~al.}(2013){Deason}, {Conroy}, {Wetzel}, \&
  {Tinker}}]{2013ApJ...777..154D}
{Deason}, A.~J., {Conroy}, C., {Wetzel}, A.~R., \& {Tinker}, J.~L. 2013, \apj,
  777, 154

\bibitem[{{Deshev} {et~al.}(2017){Deshev}, {Finoguenov}, {Verdugo}, {Ziegler},
  {Park}, {Hwang}, {Haines}, {Kamphuis}, {Tamm}, {Einasto}, {Hwang}, \&
  {Park}}]{2017arXiv170703208D}
{Deshev}, B., {Finoguenov}, A., {Verdugo}, M., {et~al.} 2017, ArXiv e-prints
  [\eprint[arXiv]{1707.03208}]

\bibitem[{{Eckert} {et~al.}(2017){Eckert}, {Gaspari}, {Owers}, {Roediger},
  {Molendi}, {Gastaldello}, {Paltani}, {Ettori}, {Venturi}, {Rossetti}, \&
  {Rudnick}}]{2017A&A...605A..25E}
{Eckert}, D., {Gaspari}, M., {Owers}, M.~S., {et~al.} 2017, \aap, 605, A25

\bibitem[{{Eckert} {et~al.}(2014){Eckert}, {Molendi}, {Owers}, {Gaspari},
  {Venturi}, {Rudnick}, {Ettori}, {Paltani}, {Gastaldello}, \&
  {Rossetti}}]{2014A&A...570A.119E}
{Eckert}, D., {Molendi}, S., {Owers}, M., {et~al.} 2014, \aap, 570, A119

\bibitem[{{Einasto} {et~al.}(2015){Einasto}, {Gramann}, {Saar},
  {Liivam{\"a}gi}, {Tempel}, {Nevalainen}, {Hein{\"a}m{\"a}ki}, {Park}, \&
  {Einasto}}]{2015A&A...580A..69E}
{Einasto}, M., {Gramann}, M., {Saar}, E., {et~al.} 2015, \aap, 580, A69

\bibitem[{{Einasto} {et~al.}(2016){Einasto}, {Lietzen}, {Gramann}, {Tempel},
  {Saar}, {Liivam{\"a}gi}, {Hein{\"a}m{\"a}ki}, {Nurmi}, \&
  {Einasto}}]{2016A&A...595A..70E}
{Einasto}, M., {Lietzen}, H., {Gramann}, M., {et~al.} 2016, \aap, 595, A70

\bibitem[{{Einasto} {et~al.}(2014){Einasto}, {Lietzen}, {Tempel}, {Gramann},
  {Liivam{\"a}gi}, \& {Einasto}}]{2014A&A...562A..87E}
{Einasto}, M., {Lietzen}, H., {Tempel}, E., {et~al.} 2014, \aap, 562, A87

\bibitem[{{Einasto} {et~al.}(2011){Einasto}, {Liivam{\"a}gi}, {Tago}, {Saar},
  {Tempel}, {Einasto}, {Mart{\'{\i}}nez}, \&
  {Hein{\"a}m{\"a}ki}}]{2011A&A...532A...5E}
{Einasto}, M., {Liivam{\"a}gi}, L.~J., {Tago}, E., {et~al.} 2011, \aap, 532, A5

\bibitem[{{Einasto} {et~al.}(2008){Einasto}, {Saar}, {Mart{\'{\i}}nez},
  {Einasto}, {Liivam{\"a}gi}, {Tago}, {Starck}, {M{\"u}ller},
  {Hein{\"a}m{\"a}ki}, {Nurmi}, {Paredes}, {Gramann}, \&
  {H{\"u}tsi}}]{2008ApJ...685...83E}
{Einasto}, M., {Saar}, E., {Mart{\'{\i}}nez}, V.~J., {et~al.} 2008, \apj, 685,
  83

\bibitem[{{Einasto} {et~al.}(2010){Einasto}, {Tago}, {Saar}, {Nurmi},
  {Enkvist}, {Einasto}, {Hein{\"a}m{\"a}ki}, {Liivam{\"a}gi}, {Tempel},
  {Einasto}, {Mart{\'{\i}}nez}, {Vennik}, \& {Pihajoki}}]{2010A&A...522A..92E}
{Einasto}, M., {Tago}, E., {Saar}, E., {et~al.} 2010, \aap, 522, A92

\bibitem[{{Einasto} {et~al.}(2012){Einasto}, {Vennik}, {Nurmi}, {Tempel},
  {Ahvensalmi}, {Tago}, {Liivam{\"a}gi}, {Saar}, {Hein{\"a}m{\"a}ki},
  {Einasto}, \& {Mart{\'{\i}}nez}}]{2012A&A...540A.123E}
{Einasto}, M., {Vennik}, J., {Nurmi}, P., {et~al.} 2012, \aap, 540, A123

\bibitem[{{Farnsworth} {et~al.}(2013){Farnsworth}, {Rudnick}, {Brown}, \&
  {Brunetti}}]{2013ApJ...779..189F}
{Farnsworth}, D., {Rudnick}, L., {Brown}, S., \& {Brunetti}, G. 2013, \apj,
  779, 189

\bibitem[{{Fo{\"e}x} {et~al.}(2017){Fo{\"e}x}, {Chon}, \&
  {B{\"o}hringer}}]{2017A&A...601A.145F}
{Fo{\"e}x}, G., {Chon}, G., \& {B{\"o}hringer}, H. 2017, \aap, 601, A145

\bibitem[{{Forman} {et~al.}(2001){Forman}, {Markevitch}, {Jones}, {Vikhlinin},
  \& {Churazov}}]{2001cghr.confE..33F}
{Forman}, W., {Markevitch}, M., {Jones}, C., {Vikhlinin}, A., \& {Churazov}, E.
  2001, in Clusters of Galaxies and the High Redshift Universe Observed in
  X-rays, ed. D.~M. {Neumann} \& J.~T.~V. {Tran}, 33

\bibitem[{{Fraley} \& {Raftery}(2006)}]{fraley2006}
{Fraley}, C. \& {Raftery}, A.~E. 2006, Technical Report, Dep. of Statistics,
  University of Washington, 504, 1

\bibitem[{{Genel} {et~al.}(2010){Genel}, {Bouch{\'e}}, {Naab}, {Sternberg}, \&
  {Genzel}}]{2010ApJ...719..229G}
{Genel}, S., {Bouch{\'e}}, N., {Naab}, T., {Sternberg}, A., \& {Genzel}, R.
  2010, \apj, 719, 229

\bibitem[{{Girardi} {et~al.}(2008){Girardi}, {Barrena}, {Boschin}, \&
  {Ellingson}}]{2008A&A...491..379G}
{Girardi}, M., {Barrena}, R., {Boschin}, W., \& {Ellingson}, E. 2008, \aap,
  491, 379

\bibitem[{{Govoni} {et~al.}(2010){Govoni}, {Dolag}, {Murgia}, {Feretti},
  {Schindler}, {Giovannini}, {Boschin}, {Vacca}, \&
  {Bonafede}}]{2010A&A...522A.105G}
{Govoni}, F., {Dolag}, K., {Murgia}, M., {et~al.} 2010, \aap, 522, A105

\bibitem[{{Gramann} {et~al.}(2015){Gramann}, {Einasto}, {Hein{\"a}m{\"a}ki},
  {Teerikorpi}, {Saar}, {Nurmi}, \& {Einasto}}]{2015A&A...581A.135G}
{Gramann}, M., {Einasto}, M., {Hein{\"a}m{\"a}ki}, P., {et~al.} 2015, \aap,
  581, A135

\bibitem[{{Haines} {et~al.}(2017{\natexlab{a}}){Haines}, {Finoguenov}, {Smith},
  {Babul}, {Egami}, {Mazzotta}, {Okabe}, {Pereira}, {Bianconi}, {McGee},
  {Ziparo}, {Campusano}, \& {Loyola}}]{2017arXiv170904945H}
{Haines}, C.~P., {Finoguenov}, A., {Smith}, G.~P., {et~al.} 2017{\natexlab{a}},
  ArXiv e-prints [\eprint[arXiv]{1709.04945}]

\bibitem[{{Haines} {et~al.}(2017{\natexlab{b}}){Haines}, {Iovino}, {Krywult},
  {Guzzo}, {Davidzon}, {Bolzonella}, {Garilli}, {Scodeggio}, {Granett}, {de la
  Torre}, {De Lucia}, {Abbas}, {Adami}, {Arnouts}, {Bottini}, {Cappi},
  {Cucciati}, {Franzetti}, {Fritz}, {Gargiulo}, {Le Brun}, {Le F{\`e}vre},
  {Maccagni}, {Ma{\l}ek}, {Marulli}, {Moutard}, {Polletta}, {Pollo}, {Tasca},
  {Tojeiro}, {Vergani}, {Zanichelli}, {Zamorani}, {Bel}, {Branchini}, {Coupon},
  {Ilbert}, {Moscardini}, {Peacock}, \& {Siudek}}]{2017A&A...605A...4H}
{Haines}, C.~P., {Iovino}, A., {Krywult}, J., {et~al.} 2017{\natexlab{b}},
  \aap, 605, A4

\bibitem[{{Haines} {et~al.}(2015){Haines}, {Pereira}, {Smith}, {Egami},
  {Babul}, {Finoguenov}, {Ziparo}, {McGee}, {Rawle}, {Okabe}, \&
  {Moran}}]{2015ApJ...806..101H}
{Haines}, C.~P., {Pereira}, M.~J., {Smith}, G.~P., {et~al.} 2015, \apj, 806,
  101

\bibitem[{{Hirv} {et~al.}(2017){Hirv}, {Pelt}, {Saar}, {Tago}, {Tamm},
  {Tempel}, \& {Einasto}}]{2017A&A...599A..31H}
{Hirv}, A., {Pelt}, J., {Saar}, E., {et~al.} 2017, \aap, 599, A31

\bibitem[{{Hopkins} {et~al.}(2005){Hopkins}, {Bahcall}, \&
  {Bode}}]{2005ApJ...618....1H}
{Hopkins}, P.~F., {Bahcall}, N.~A., \& {Bode}, P. 2005, \apj, 618, 1

\bibitem[{{Huchra} \& {Geller}(1982)}]{1982ApJ...257..423H}
{Huchra}, J.~P. \& {Geller}, M.~J. 1982, \apj, 257, 423

\bibitem[{{Ihaka} \& {Gentleman}(1996)}]{ig96}
{Ihaka}, R. \& {Gentleman}, R. 1996, Journal of Computational and Graphical
  Statistics, 5, 299

\bibitem[{{J{\~o}eveer} \& {Einasto}(1978)}]{Joeveer:1978a}
{J{\~o}eveer}, M. \& {Einasto}, J. 1978, in IAU Symposium, Vol.~79, Large Scale
  Structures in the Universe, ed. M.~S. {Longair} \& J.~{Einasto}, 241--250

\bibitem[{{J{\~o}eveer} {et~al.}(1978){J{\~o}eveer}, {Einasto}, \&
  {Tago}}]{1978MNRAS.185..357J}
{J{\~o}eveer}, M., {Einasto}, J., \& {Tago}, E. 1978, \mnras, 185, 357

\bibitem[{{Jaff{\'e}} {et~al.}(2015){Jaff{\'e}}, {Smith}, {Candlish},
  {Poggianti}, {Sheen}, \& {Verheijen}}]{2015MNRAS.448.1715J}
{Jaff{\'e}}, Y.~L., {Smith}, R., {Candlish}, G.~N., {et~al.} 2015, \mnras, 448,
  1715

\bibitem[{{Johnston-Hollitt} {et~al.}(2008){Johnston-Hollitt}, {Sato}, {Gill},
  {Fleenor}, \& {Brick}}]{2008MNRAS.390..289J}
{Johnston-Hollitt}, M., {Sato}, M., {Gill}, J.~A., {Fleenor}, M.~C., \&
  {Brick}, A.-M. 2008, \mnras, 390, 289

\bibitem[{{Kauffmann} {et~al.}(2003{\natexlab{a}}){Kauffmann}, {Heckman},
  {White}, {Charlot}, {Tremonti}, {Brinchmann}, {Bruzual}, {Peng}, {Seibert},
  {Bernardi}, {Blanton}, {Brinkmann}, {Castander}, {Cs{\'a}bai}, {Fukugita},
  {Ivezic}, {Munn}, {Nichol}, {Padmanabhan}, {Thakar}, {Weinberg}, \&
  {York}}]{2003MNRAS.341...33K}
{Kauffmann}, G., {Heckman}, T.~M., {White}, S.~D.~M., {et~al.}
  2003{\natexlab{a}}, \mnras, 341, 33

\bibitem[{{Kauffmann} {et~al.}(2003{\natexlab{b}}){Kauffmann}, {Heckman},
  {White}, {Charlot}, {Tremonti}, {Peng}, {Seibert}, {Brinkmann}, {Nichol},
  {SubbaRao}, \& {York}}]{2003MNRAS.341...54K}
{Kauffmann}, G., {Heckman}, T.~M., {White}, S.~D.~M., {et~al.}
  2003{\natexlab{b}}, \mnras, 341, 54

\bibitem[{{Kim} {et~al.}(2015){Kim}, {Park}, {L'Huillier}, \&
  {Hong}}]{2015JKAS...48..213K}
{Kim}, J., {Park}, C., {L'Huillier}, B., \& {Hong}, S.~E. 2015, Journal of
  Korean Astronomical Society, 48, 213

\bibitem[{{Kofman} \& {Shandarin}(1988)}]{1988Natur.334..129K}
{Kofman}, L.~A. \& {Shandarin}, S.~F. 1988, \nat, 334, 129

\bibitem[{{Komatsu} {et~al.}(2011){Komatsu}, {Smith}, {Dunkley}, {Bennett},
  {Gold}, {Hinshaw}, {Jarosik}, {Larson}, {Nolta}, {Page}, {Spergel},
  {Halpern}, {Hill}, {Kogut}, {Limon}, {Meyer}, {Odegard}, {Tucker}, {Weiland},
  {Wollack}, \& {Wright}}]{2011ApJS..192...18K}
{Komatsu}, E., {Smith}, K.~M., {Dunkley}, J., {et~al.} 2011, \apjs, 192, 18

\bibitem[{{Krause} {et~al.}(2013){Krause}, {Ribeiro}, \&
  {Lopes}}]{2013A&A...551A.143K}
{Krause}, M.~O., {Ribeiro}, A.~L.~B., \& {Lopes}, P.~A.~A. 2013, \aap, 551,
  A143

\bibitem[{{Kravtsov} \& {Borgani}(2012)}]{2012ARA&A..50..353K}
{Kravtsov}, A.~V. \& {Borgani}, S. 2012, \araa, 50, 353

\bibitem[{{Lietzen} {et~al.}(2012){Lietzen}, {Tempel}, {Hein{\"a}m{\"a}ki},
  {Nurmi}, {Einasto}, \& {Saar}}]{2012A&A...545A.104L}
{Lietzen}, H., {Tempel}, E., {Hein{\"a}m{\"a}ki}, P., {et~al.} 2012, \aap, 545,
  A104

\bibitem[{{Liivam{\"a}gi} {et~al.}(2012){Liivam{\"a}gi}, {Tempel}, \&
  {Saar}}]{2012A&A...539A..80L}
{Liivam{\"a}gi}, L.~J., {Tempel}, E., \& {Saar}, E. 2012, \aap, 539, A80

\bibitem[{{Liu} {et~al.}(2016){Liu}, {Yu}, {Tozzi}, \&
  {Zhu}}]{2016ApJ...821...29L}
{Liu}, A., {Yu}, H., {Tozzi}, P., \& {Zhu}, Z.-H. 2016, \apj, 821, 29

\bibitem[{{Luparello} {et~al.}(2013){Luparello}, {Lares}, {Yaryura}, {Paz},
  {Padilla}, \& {Lambas}}]{2013MNRAS.432.1367L}
{Luparello}, H.~E., {Lares}, M., {Yaryura}, C.~Y., {et~al.} 2013, \mnras, 432,
  1367

\bibitem[{{Mahajan} {et~al.}(2012){Mahajan}, {Raychaudhury}, \&
  {Pimbblet}}]{2012MNRAS.427.1252M}
{Mahajan}, S., {Raychaudhury}, S., \& {Pimbblet}, K.~A. 2012, \mnras, 427, 1252

\bibitem[{{Mahdavi} {et~al.}(2007){Mahdavi}, {Hoekstra}, {Babul}, {Balam}, \&
  {Capak}}]{2007ApJ...668..806M}
{Mahdavi}, A., {Hoekstra}, H., {Babul}, A., {Balam}, D.~D., \& {Capak}, P.~L.
  2007, \apj, 668, 806

\bibitem[{{Maraston} {et~al.}(2009){Maraston}, {Str{\"o}mb{\"a}ck}, {Thomas},
  {Wake}, \& {Nichol}}]{2009MNRAS.394L.107M}
{Maraston}, C., {Str{\"o}mb{\"a}ck}, G., {Thomas}, D., {Wake}, D.~A., \&
  {Nichol}, R.~C. 2009, \mnras, 394, L107

\bibitem[{{Markevitch} {et~al.}(2000){Markevitch}, {Ponman}, {Nulsen}, {Bautz},
  {Burke}, {David}, {Davis}, {Donnelly}, {Forman}, {Jones}, {Kaastra},
  {Kellogg}, {Kim}, {Kolodziejczak}, {Mazzotta}, {Pagliaro}, {Patel}, {Van
  Speybroeck}, {Vikhlinin}, {Vrtilek}, {Wise}, \& {Zhao}}]{2000ApJ...541..542M}
{Markevitch}, M., {Ponman}, T.~J., {Nulsen}, P.~E.~J., {et~al.} 2000, \apj,
  541, 542

\bibitem[{{Martel} {et~al.}(2014){Martel}, {Robichaud}, \&
  {Barai}}]{2014ApJ...786...79M}
{Martel}, H., {Robichaud}, F., \& {Barai}, P. 2014, \apj, 786, 79

\bibitem[{{Mart{\'{\i}}nez} \& {Saar}(2002)}]{2002sgd..book.....M}
{Mart{\'{\i}}nez}, V.~J. \& {Saar}, E. 2002, {Statistics of the Galaxy
  Distribution} (Chapman {\&} Hall/CRC, Boca Raton)

\bibitem[{{McIntosh} {et~al.}(2016){McIntosh}, {de Propris}, \&
  {West}}]{2016AAS...22723518M}
{McIntosh}, M., {de Propris}, R., \& {West}, M. 2016, in American Astronomical
  Society Meeting Abstracts, Vol. 227, American Astronomical Society Meeting
  Abstracts, 235.18

\bibitem[{{Merluzzi} {et~al.}(2015){Merluzzi}, {Busarello}, {Haines},
  {Mercurio}, {Okabe}, {Pimbblet}, {Dopita}, {Grado}, {Limatola}, {Bourdin},
  {Mazzotta}, {Capaccioli}, {Napolitano}, \& {Schipani}}]{2015MNRAS.446..803M}
{Merluzzi}, P., {Busarello}, G., {Haines}, C.~P., {et~al.} 2015, \mnras, 446,
  803

\bibitem[{{Molnar} {et~al.}(2009){Molnar}, {Hearn}, {Haiman}, {Bryan},
  {Evrard}, \& {Lake}}]{2009ApJ...696.1640M}
{Molnar}, S.~M., {Hearn}, N., {Haiman}, Z., {et~al.} 2009, \apj, 696, 1640

\bibitem[{{Moster} {et~al.}(2017){Moster}, {Naab}, \&
  {White}}]{2017arXiv170505373M}
{Moster}, B.~P., {Naab}, T., \& {White}, S.~D.~M. 2017, ArXiv e-prints
  [\eprint[arXiv]{1705.05373}]

\bibitem[{{Moster} {et~al.}(2010){Moster}, {Somerville}, {Maulbetsch}, {van den
  Bosch}, {Macci{\`o}}, {Naab}, \& {Oser}}]{2010ApJ...710..903M}
{Moster}, B.~P., {Somerville}, R.~S., {Maulbetsch}, C., {et~al.} 2010, \apj,
  710, 903

\bibitem[{{Mulroy} {et~al.}(2017){Mulroy}, {McGee}, {Gillman}, {Smith},
  {Haines}, {D{\'e}mocl{\`e}s}, {Okabe}, \& {Egami}}]{2017MNRAS.472.3246M}
{Mulroy}, S.~L., {McGee}, S.~L., {Gillman}, S., {et~al.} 2017, \mnras, 472,
  3246

\bibitem[{{Munari} {et~al.}(2014){Munari}, {Biviano}, \&
  {Mamon}}]{2014A&A...566A..68M}
{Munari}, E., {Biviano}, A., \& {Mamon}, G.~A. 2014, \aap, 566, A68

\bibitem[{{Munari} {et~al.}(2016){Munari}, {Grillo}, {De Lucia}, {Biviano},
  {Annunziatella}, {Borgani}, {Lombardi}, {Mercurio}, \&
  {Rosati}}]{2016ApJ...827L...5M}
{Munari}, E., {Grillo}, C., {De Lucia}, G., {et~al.} 2016, \apjl, 827, L5

\bibitem[{{Muzzin} {et~al.}(2014){Muzzin}, {van der Burg}, {McGee}, {Balogh},
  {Franx}, {Hoekstra}, {Hudson}, {Noble}, {Taranu}, {Webb}, {Wilson}, \&
  {Yee}}]{2014ApJ...796...65M}
{Muzzin}, A., {van der Burg}, R.~F.~J., {McGee}, S.~L., {et~al.} 2014, \apj,
  796, 65

\bibitem[{{Okabe} \& {Umetsu}(2008)}]{2008PASJ...60..345O}
{Okabe}, N. \& {Umetsu}, K. 2008, \pasj, 60, 345

\bibitem[{{Old} {et~al.}(2017){Old}, {Wojtak}, {Pearce}, {Gray}, {Mamon},
  {Sif{\'o}n}, {Tempel}, {Biviano}, {Yee}, {de Carvalho}, {M{\"u}ller}, {Sepp},
  {Skibba}, {Croton}, {Power}, {von der Linden}, \&
  {Saro}}]{2017arXiv170910108O}
{Old}, L., {Wojtak}, R., {Pearce}, F.~R., {et~al.} 2017, ArXiv e-prints
  [\eprint[arXiv]{1709.10108}]

\bibitem[{{Oman} {et~al.}(2013){Oman}, {Hudson}, \&
  {Behroozi}}]{2013MNRAS.431.2307O}
{Oman}, K.~A., {Hudson}, M.~J., \& {Behroozi}, P.~S. 2013, \mnras, 431, 2307

\bibitem[{{O'Mill} {et~al.}(2015){O'Mill}, {Proust}, {Capelato}, {Castejon},
  {Cypriano}, {Neto}, \& {Laerte}}]{2015MNRAS.453..868O}
{O'Mill}, A.~L., {Proust}, D., {Capelato}, H.~V., {et~al.} 2015, \mnras, 453,
  868

\bibitem[{{Owers} {et~al.}(2011){Owers}, {Nulsen}, \&
  {Couch}}]{2011ApJ...741..122O}
{Owers}, M.~S., {Nulsen}, P.~E.~J., \& {Couch}, W.~J. 2011, \apj, 741, 122

\bibitem[{{Paccagnella} {et~al.}(2017){Paccagnella}, {Vulcani}, {Poggianti},
  {Fritz}, {Fasano}, {Moretti}, {Jaff{\'e}}, {Biviano}, {Gullieuszik},
  {Bettoni}, {Cava}, {Couch}, \& {D'Onofrio}}]{2017ApJ...838..148P}
{Paccagnella}, A., {Vulcani}, B., {Poggianti}, B.~M., {et~al.} 2017, \apj, 838,
  148

\bibitem[{{Paccagnella} {et~al.}(2016){Paccagnella}, {Vulcani}, {Poggianti},
  {Moretti}, {Fritz}, {Gullieuszik}, {Couch}, {Bettoni}, {Cava}, {D'Onofrio},
  \& {Fasano}}]{2016ApJ...816L..25P}
{Paccagnella}, A., {Vulcani}, B., {Poggianti}, B.~M., {et~al.} 2016, \apjl,
  816, L25

\bibitem[{{Park} \& {Hwang}(2009)}]{2009ApJ...699.1595P}
{Park}, C. \& {Hwang}, H.~S. 2009, \apj, 699, 1595

\bibitem[{{Paz} {et~al.}(2011){Paz}, {Sgr{\'o}}, {Merch{\'a}n}, \&
  {Padilla}}]{2011MNRAS.414.2029P}
{Paz}, D.~J., {Sgr{\'o}}, M.~A., {Merch{\'a}n}, M., \& {Padilla}, N. 2011,
  \mnras, 414, 2029

\bibitem[{{Pearson} {et~al.}(2014){Pearson}, {Batiste}, \&
  {Batuski}}]{2014MNRAS.441.1601P}
{Pearson}, D.~W., {Batiste}, M., \& {Batuski}, D.~J. 2014, \mnras, 441, 1601

\bibitem[{{Plionis} \& {Basilakos}(2002)}]{2002MNRAS.329L..47P}
{Plionis}, M. \& {Basilakos}, S. 2002, \mnras, 329, L47

\bibitem[{{Porter} {et~al.}(2008){Porter}, {Raychaudhury}, {Pimbblet}, \&
  {Drinkwater}}]{2008MNRAS.388.1152P}
{Porter}, S.~C., {Raychaudhury}, S., {Pimbblet}, K.~A., \& {Drinkwater}, M.~J.
  2008, \mnras, 388, 1152

\bibitem[{{Proust} {et~al.}(2006){Proust}, {Quintana}, {Carrasco},
  {Reisenegger}, {Slezak}, {Muriel}, {D{\"u}nner}, {Sodr{\'e}}, {Drinkwater},
  {Parker}, \& {Ragone}}]{2006A&A...447..133P}
{Proust}, D., {Quintana}, H., {Carrasco}, E.~R., {et~al.} 2006, \aap, 447, 133

\bibitem[{{Reisenegger} {et~al.}(2000){Reisenegger}, {Quintana}, {Carrasco}, \&
  {Maze}}]{2000AJ....120..523R}
{Reisenegger}, A., {Quintana}, H., {Carrasco}, E.~R., \& {Maze}, J. 2000, \aj,
  120, 523

\bibitem[{{Rhee} {et~al.}(2017){Rhee}, {Smith}, {Choi}, {Yi}, {Jaff{\'e}},
  {Candlish}, \& {S{\'a}nchez-J{\'a}nssen}}]{2017ApJ...843..128R}
{Rhee}, J., {Smith}, R., {Choi}, H., {et~al.} 2017, \apj, 843, 128

\bibitem[{{Rossetti} {et~al.}(2013){Rossetti}, {Eckert}, {De Grandi},
  {Gastaldello}, {Ghizzardi}, {Roediger}, \& {Molendi}}]{2013A&A...556A..44R}
{Rossetti}, M., {Eckert}, D., {De Grandi}, S., {et~al.} 2013, \aap, 556, A44

\bibitem[{{Salim} {et~al.}(2007){Salim}, {Rich}, {Charlot}, {Brinchmann},
  {Johnson}, {Schiminovich}, {Seibert}, {Mallery}, {Heckman}, {Forster},
  {Friedman}, {Martin}, {Morrissey}, {Neff}, {Small}, {Wyder}, {Bianchi},
  {Donas}, {Lee}, {Madore}, {Milliard}, {Szalay}, {Welsh}, \&
  {Yi}}]{2007ApJS..173..267S}
{Salim}, S., {Rich}, R.~M., {Charlot}, S., {et~al.} 2007, \apjs, 173, 267

\bibitem[{{Sarzi} {et~al.}(2006){Sarzi}, {Falc{\'o}n-Barroso}, {Davies},
  {Bacon}, {Bureau}, {Cappellari}, {de Zeeuw}, {Emsellem}, {Fathi},
  {Krajnovi{\'c}}, {Kuntschner}, {McDermid}, \&
  {Peletier}}]{2006MNRAS.366.1151S}
{Sarzi}, M., {Falc{\'o}n-Barroso}, J., {Davies}, R.~L., {et~al.} 2006, \mnras,
  366, 1151

\bibitem[{{Shankar} {et~al.}(2009){Shankar}, {Bernardi}, \&
  {Haiman}}]{2009ApJ...694..867S}
{Shankar}, F., {Bernardi}, M., \& {Haiman}, Z. 2009, \apj, 694, 867

\bibitem[{{Small} {et~al.}(1998){Small}, {Ma}, {Sargent}, \&
  {Hamilton}}]{1998ApJ...492...45S}
{Small}, T.~A., {Ma}, C.-P., {Sargent}, W.~L.~W., \& {Hamilton}, D. 1998, \apj,
  492, 45

\bibitem[{{Song} {et~al.}(2017){Song}, {Hwang}, {Park}, \&
  {Tamura}}]{2017ApJ...842...88S}
{Song}, H., {Hwang}, H.~S., {Park}, C., \& {Tamura}, T. 2017, \apj, 842, 88

\bibitem[{{Suhhonenko} {et~al.}(2011){Suhhonenko}, {Einasto}, {Liivam{\"a}gi},
  {Saar}, {Einasto}, {H{\"u}tsi}, {M{\"u}ller}, {Starobinsky}, {Tago}, \&
  {Tempel}}]{2011A&A...531A.149S}
{Suhhonenko}, I., {Einasto}, J., {Liivam{\"a}gi}, L.~J., {et~al.} 2011, \aap,
  531, A149

\bibitem[{{Tchernin} {et~al.}(2016){Tchernin}, {Eckert}, {Ettori},
  {Pointecouteau}, {Paltani}, {Molendi}, {Hurier}, {Gastaldello}, {Lau},
  {Nagai}, {Roncarelli}, \& {Rossetti}}]{2016A&A...595A..42T}
{Tchernin}, C., {Eckert}, D., {Ettori}, S., {et~al.} 2016, \aap, 595, A42

\bibitem[{{Tempel} {et~al.}(2015){Tempel}, {Guo}, {Kipper}, \&
  {Libeskind}}]{2015MNRAS.450.2727T}
{Tempel}, E., {Guo}, Q., {Kipper}, R., \& {Libeskind}, N.~I. 2015, \mnras, 450,
  2727

\bibitem[{{Tempel} {et~al.}(2016){Tempel}, {Stoica}, {Kipper}, \&
  {Saar}}]{2016A&C....16...17T}
{Tempel}, E., {Stoica}, R.~S., {Kipper}, R., \& {Saar}, E. 2016, Astronomy and
  Computing, 16, 17

\bibitem[{{Tempel} {et~al.}(2014{\natexlab{a}}){Tempel}, {Stoica},
  {Mart{\'{\i}}nez}, {Liivam{\"a}gi}, {Castellan}, \&
  {Saar}}]{2014MNRAS.438.3465T}
{Tempel}, E., {Stoica}, R.~S., {Mart{\'{\i}}nez}, V.~J., {et~al.}
  2014{\natexlab{a}}, \mnras, 438, 3465

\bibitem[{{Tempel} {et~al.}(2012){Tempel}, {Tago}, \&
  {Liivam{\"a}gi}}]{2012A&A...540A.106T}
{Tempel}, E., {Tago}, E., \& {Liivam{\"a}gi}, L.~J. 2012, \aap, 540, A106

\bibitem[{{Tempel} \& {Tamm}(2015)}]{2015A&A...576L...5T}
{Tempel}, E. \& {Tamm}, A. 2015, \aap, 576, L5

\bibitem[{{Tempel} {et~al.}(2014{\natexlab{b}}){Tempel}, {Tamm}, {Gramann},
  {Tuvikene}, {Liivam{\"a}gi}, {Suhhonenko}, {Kipper}, {Einasto}, \&
  {Saar}}]{2014A&A...566A...1T}
{Tempel}, E., {Tamm}, A., {Gramann}, M., {et~al.} 2014{\natexlab{b}}, \aap,
  566, A1

\bibitem[{{Tremonti} {et~al.}(2004){Tremonti}, {Heckman}, {Kauffmann},
  {Brinchmann}, {Charlot}, {White}, {Seibert}, {Peng}, {Schlegel}, {Uomoto},
  {Fukugita}, \& {Brinkmann}}]{2004ApJ...613..898T}
{Tremonti}, C.~A., {Heckman}, T.~M., {Kauffmann}, G., {et~al.} 2004, \apj, 613,
  898

\bibitem[{{van de Weygaert} \& {Schaap}(2009)}]{2009LNP...665..291V}
{van de Weygaert}, R. \& {Schaap}, W. 2009, in Lecture Notes in Physics, Berlin
  Springer Verlag, Vol. 665, Data Analysis in Cosmology, ed. V.~J.
  {Mart{\'{\i}}nez}, E.~{Saar}, E.~{Mart{\'{\i}}nez-Gonz{\'a}lez}, \& M.-J.
  {Pons-Border{\'{\i}}a}, 291--413

\bibitem[{{Venturi} {et~al.}(2017){Venturi}, {Rossetti}, {Brunetti},
  {Farnsworth}, {Gastaldello}, {Giacintucci}, {Lal}, {Rudnick}, {Shimwell},
  {Eckert}, {Molendi}, \& {Owers}}]{2017A&A...603A.125V}
{Venturi}, T., {Rossetti}, M., {Brunetti}, G., {et~al.} 2017, \aap, 603, A125

\bibitem[{{Vijayaraghavan} \& {Ricker}(2013)}]{2013MNRAS.435.2713V}
{Vijayaraghavan}, R. \& {Ricker}, P.~M. 2013, \mnras, 435, 2713

\bibitem[{{Weinzirl} {et~al.}(2017){Weinzirl}, {Arag{\'o}n-Salamanca}, {Gray},
  {Bamford}, {Rodr{\'{\i}}guez del Pino}, {Chies-Santos}, {B{\"o}hm}, {Wolf},
  \& {Cool}}]{2017MNRAS.471..182W}
{Weinzirl}, T., {Arag{\'o}n-Salamanca}, A., {Gray}, M.~E., {et~al.} 2017,
  \mnras, 471, 182

\bibitem[{{West} \& {Blakeslee}(2000)}]{2000ApJ...543L..27W}
{West}, M.~J. \& {Blakeslee}, J.~P. 2000, \apjl, 543, L27

\bibitem[{{West} {et~al.}(2017){West}, {de Propris}, {Bremer}, \&
  {Phillipps}}]{2017NatAs...1E.157W}
{West}, M.~J., {de Propris}, R., {Bremer}, M.~N., \& {Phillipps}, S. 2017,
  Nature Astronomy, 1, 0157

\bibitem[{{Wittman} {et~al.}(2017){Wittman}, {Golovich}, \&
  {Dawson}}]{2017arXiv170105877W}
{Wittman}, D., {Golovich}, N., \& {Dawson}, W.~A. 2017, ArXiv e-prints
  [\eprint[arXiv]{1701.05877}]

\bibitem[{{Yoon} {et~al.}(2017){Yoon}, {Chung}, {Smith}, \&
  {Jaff{\'e}}}]{2017ApJ...838...81Y}
{Yoon}, H., {Chung}, A., {Smith}, R., \& {Jaff{\'e}}, Y.~L. 2017, \apj, 838, 81

\bibitem[{{Zeldovich} {et~al.}(1982){Zeldovich}, {Einasto}, \&
  {Shandarin}}]{1982Natur.300..407Z}
{Zeldovich}, I.~B., {Einasto}, J., \& {Shandarin}, S.~F. 1982, \nat, 300, 407

\end{thebibliography}

\end{document}